\documentclass[10pt]{article}

\usepackage[font=small,labelfont=bf]{caption}
\usepackage{graphicx}
\usepackage{amsmath}
\usepackage{bm}
\usepackage{amsfonts}
\usepackage{amssymb}
\usepackage{mathtools}
\usepackage{mathrsfs}
\usepackage{multirow}
\usepackage{url}
\usepackage{hyperref}
\usepackage{algorithm}
\usepackage{algpseudocode}
\usepackage[draft]{fixme}
\usepackage{relsize}
\usepackage{authblk}
\usepackage{caption}
\usepackage{subcaption}
\usepackage{framed}
\usepackage{booktabs} 
\usepackage{pgfplots, pgfplotstable}
\usepackage[normalem]{ulem}
\usepackage{comment}
\usepackage{tikz}
\usepackage{geometry}
\newcommand\norm[1]{\lVert#1\rVert}

\makeatother

\newcommand*{\figref}[2][]{%
  \hyperref[{#2}]{%
    \ref*{#2}%
    \ifx\\#1\\%
    \else
      #1%
    \fi
  }%
}

\pgfplotstableset{
  every last row/.style={after row=\bottomrule},
}

\begin{document}

\title{Examining the simulation-to-reality gap of a wheel loader digging in deformable terrain}

\author[1,2]{Koji Aoshima}
\author[2,3]{Martin Servin}
\affil[1]{Komatsu Ltd.}
\affil[2]{Umeå University}
\affil[3]{Algoryx Simulation}

\date{\today}

\maketitle

\begin{abstract}
We investigate how well a physics-based simulator can replicate a real wheel loader performing bucket filling in a pile of soil. The comparison is made using 
field test time series of the vehicle motion and actuation forces, loaded mass, and total work.
The vehicle was modeled as a rigid multibody system with frictional contacts, driveline, and 
linear actuators. For the soil, we tested discrete element models of different resolutions, 
with and without multiscale acceleration. The spatio-temporal resolution ranged between 50\--400 mm and 2-500 ms, and the computational speed 
was between 1/10,000 to 5 times faster than real-time.
The simulation-to-reality gap was found to be around 10\% and exhibited a weak dependence on the level of fidelity, 
e.g., compatible with real-time simulation. Furthermore, the sensitivity of an optimized force feedback 
controller under transfer between different simulation domains was investigated. 
The domain bias was observed to cause a performance reduction of 5\% despite
the domain gap being about 15\%.
\end{abstract}

\section{Introduction}
\label{sec:introduction}
Simulators are essential for developing autonomous control of heavy equipment and rough-terrain vehicles. 
They offer a safe and efficient way to conduct controlled and repeatable experiments for testing and optimizing 
the performance in the early development stages. 
This makes it possible to generate large amounts of annotated synthetic 
training data needed for leveraging deep learning methods \cite{Matsumoto2020,Kurinov2020,backman2021,wiberg:2021:crt,Egli2022b}.

Limiting factors are the computational speed and how 
accurately the simulator reflects the real system \cite{choi2021use}. 
Having a reality gap is unavoidable, but when the discrepancy between the simulated and real system is too large, 
a solution optimized in the simulated domain transfers poorly to the real domain \cite{koos2010,wiberg2023}. On the
other hand, a finely resolved simulator easily becomes too slow to run the simulations needed
to distinguish between near-optimal, low-performing, or hazardous solutions. 

For earthmoving equipment, there is little knowledge about how the reality gap should be measured, how it depends on
the simulator's level of resolution, or what effect it has on the transferability of the results. To this end,
we construct wheel loader simulators of different levels of fidelity and examine how they differ from each other
and from a real wheel loader performing bucket-filling operations. The comparison is made through the lens of 
synthetic and real sensor data that may be used for automatic bucket filling with force feedback control.

Two types of simulators are used. In both cases, the vehicle is modeled as a 3D rigid multibody system
with frictional contacts and nonsmooth dynamics. The difference lies in that the terrain is resolved using either a discrete
element model (DEM) or modeled using a reduced multiscale model. 
The latter can be understood as a multibody dynamics generalization of the fundamental earth moving equation (FEE) that is conventionally used in vehicle-soil mechanics analysis but is limited to stationary conditions. The general idea is to predict a zone of active soil deformations and represent this as a dynamic body and additional multibody constraint added to the vehicle system. The mass flow in the active zone is approximated with a co-simulated DEM model. There exist several realizations of the general idea in several physics engines \cite{Holz2009,Jaiswal2019,Servin2021}.
																	
The simulators' spatio-temporal
resolution ranges between 50\--400 mm and 2\--500 ms, with computational speed 
between $10^{-4}$ and $5$ times faster than real-time.  The simulators are equipped with the
same sensing capabilities used in the field tests, which include kinematics and force
sensors in selected joints and actuators, weight estimation of the loaded material, and the shape of the pile surface before
loading. The simulators and field tests are compared using the measured time series,
loaded mass, and mechanical work. The operator’s control of the vehicle is replicated using feedforward control. 
Finally, we investigate the domain sensitivity of a force feedback controller optimized
in a real-time simulator under transfer to a simulator of much higher fidelity.

\section{Related work}
\label{sec:related_work}
 
In the scientific literature, there are few examples of full system simulators that represent 
the full dynamics of both a wheel loader and its environment. Exceptions include studies for 
predicting \cite{aoshima2021} and optimizing \cite{Lindmark2018} the outcome of a 
dig plan given a soil pile of a certain shape, and loader automation using deep reinforcement 
learning-based control \cite{backman2021} or nonlinear model 
predictive control \cite{song2022}. Only in \cite{aoshima2021} were the simulators directly 
compared to field tests with a real wheel loader. The present study is a direct extension of
this work. 

In \cite{Azulay2021}, a controller for a wheeled scooping robot was developed 
using deep reinforcement learning in a simulated environment and then transferred to a physical 
robot without any domain adaptation. Unsurprisingly, notable differences were observed between the simulated and real
bucket trajectories, scooped mass, and loading time. No conclusion was made about what kind 
of discrepancy between the simulated and real systems caused the difference in outcome.
A high sensitivity to changing contact forces was reported. A possible explanation is that the
simulator used too coarse particles. The bucket could occupy roughly 15 particles, while
the real material was much more fine-grained.

There are several simulation studies of the relationship between bucket trajectory, fill factor, and mechanical work
using the discrete element method (DEM) for the soil and a kinematically 
controlled bucket geometry \cite{Filla2017,Frank2018,meng2019,Wang2022a}. However, as pointed out in the review article 
\cite{Dadhich2016}, it is generally impossible to track precisely a prescribed dig trajectory
because of the unpredictive nature of the soil-vehicle interaction, e.g., 
soil flow and wheel slip.  Worse yet, a kinematically feasible trajectory might not be realizable 
with the soil dynamics and physical limitations of driveline and hydraulic actuation at hand.

In \cite{Frank2018}, DEM simulations in quasi-2D were carried out with a kinematically controlled bucket
loading gravel along numerous pre-planned trajectories. The bucket velocity and force were input
to a mapping function that outputs the corresponding velocities and forces in the 
lift and tilt cylinders. The optimal trajectory and control were found
through dynamic programming using a wheel loader model that included the engine, driveline, and hydraulics.
The optimal solution had 14\% higher fuel consumption than the most fuel-efficient loading cycle
among the field tests carried out with skilled operators.
The optimization assumes no simulation-to-reality gap while devising a scheme that avoids additional system simulations,
at the time, requiring 23 CPU hours for each loading cycle.

In \cite{Wang2022a}, time series measurements from field tests were fed into a kinematically controlled loader mechanism 
and co-simulated with a DEM representation of the soil. The simulated and real 
working resistance agreed with an average deviation of 7\%,
however, with no explanation of how the pile shape and DEM model parameters were set.

There is extensive literature on models for the wheel loader dynamics alone or coupled
with simplified models for the force on the bucket from the soil. One elaborate model, including the dynamics of
the articulated multibody system, hydromechanical powertrain, and tires, 
was developed and validated in \cite{Oh2015,Kim2016}, for the purpose of analysis and optimization 
of working patterns and energy flow for various working cycles. The empirical material model 
predicts forces on the bucket but does not explicitly model the soil dynamics and was not
evaluated on this.

\section{The simulation-to-reality gap}
\label{sec:sim2realGap}

A simulator is an idealized replica of a real system, and it is unavoidable that it
behaves somewhat differently, although fed with identical control signals. The potential causes
for the mismatch can broadly be categorized into model errors, numerical errors, and 
implementation errors.
Model errors include unmodeled or oversimplified geometry and physics, inaccurate model 
parameters and initial conditions. When the system involves 
feedback control, actuator latency and noise are reportedly major sources of model errors 
\cite{ibarz2021}.
Common sources for numerical errors are using a spatial and temporal resolution that is too coarse.
Multiphysics and multiscale simulations are prone to solver and co-simulation coupling errors. 
When simulations run over a long time, compared to the integration timestep, it is important
to use numerically stable algorithms that prevent locally small errors from accumulating
into large global errors. Low-order variational integrators, preserving the fundamental symmetries,
are then advantageous over the standard Runge-Kutta or multistep methods with high local accuracy but without
global error bounds \cite{hairer:1996:sod}.
Machine learning algorithms that rely on system state exploration, e.g., reinforcement learning, are particularly sensitive to 
simulator imperfections. RL agents are prone to exploit simulation errors 
if there is an advantage to it. An illustrative example is the use of unphysical collision dynamics in \cite{Kadian2020}, triggered by sliding along walls into the corners to find shortcuts through otherwise non-navigable space. 

In the field of robotics and deep learning, the discrepancy between a simulated and real 
system is usually referred to as the \emph{reality gap}, \emph{simulation-to-reality gap}, or 
\emph{sim-to-real gap} in short \cite{zhao2020sim2real}. If the gap is significant,
a solution developed in simulation will exhibit a \emph{simulation bias} and cause it
to perform differently, and usually poorly, when transferred to the real system \cite{atkeson1997}.
The gap is severe if the effort to adapt the solution to the real domain is greater than
its conception in the simulated environment.  The reality gap may be considered small when it is 
less than the natural variations in different instances of the real system.
Hence, there is no objective measure for the sim-to-real gap.
It depends on the task the system intends to perform and is relative to the naturally
occurring variations. 

System identification is the process of optimizing the model parameters, $\bm{\theta}$, given a 
measure of the discrepancy between the simulated and real behavior. The classical techniques
of frequency and impulse response methods focus on linear systems such that the best parameter fit 
results in a linear least square problem \cite{soderstrom1989system}.  In \cite{tan2016}, 
the system identification is stated in terms of the average trajectory deviation using an Euclidean weighted norm  
\begin{equation}\label{eq:sysID}
  \bm{\theta} = \arg\min \frac{1}{k} \sum_{i=1}^k \int_0^{T}\norm{\bm{y}_i(t;\bm{\theta}) - \hat{\bm{y}}_i(t)}^2_W \text{d}t,
\end{equation}
																 
where $\bm{y}_i(t;\bm{\theta})$ and $\hat{\bm{y}}_i(t)$ are the simulated and real trajectories,
respectively. The average is over a sequence of $k$ reference trajectories.
The system state vector $\bm{y}$ may be represented in either reduced or maximal
coordinates and a weight matrix $W$ for controlling the 
relative importance of each degree of freedom. The gradient-free Covariance Matrix
Adaptation\footnote{CMA is a 
stochastic sampling-based optimization algorithm, which has
been successfully applied to search for control parameters
when the problem domain is highly discontinuous.} (CMA) was used as optimizer  
due to the presence of intermittent contacts and the complex interplay between the 
simulation results and the simulation parameters.

The need for metrics and benchmark data for the sim-to-real gap was recognized in \cite{collins2019}. 
Benchmark data was collected for ten different robotic manipulation tasks
using a motion capture system for the pose and robotic force/torque sensors. 
The metrics included the Euclidean distance error of real and simulated end effector position, 
rotation, and pose as the distance on the Euclidean group SE(3) that combines translation and rotation, 
velocity, acceleration, motor torque, and contact-induced force and moment. When the task involves 
manipulating a (rigid) object, velocity and acceleration error measures for this were included as well.

The sim-to-real gap in the context of robotic manipulation is often attributed
to frictional contact modeling and solvers \cite{muratore2022,Horak2019}.  Direct solvers
usually rely on linearization of the friction law using a box or polyhedral
discretization of the Coulomb cone. This may induce an artificial directional
dependency. Iterative solvers leave a truncation error that often appears as numerical
elasticity and damping, and excessive sliding. One way to evaluate numerical errors
is the \emph{self-consistency error} \cite{erez2015}, i.e., comparing a numerical 
solution with a reference solution computed with the same model and simulator but with
finest possible setting for spatial and temporal resolution and solver settings.
The error should be interpreted carefully. A small self-consistency error does not guarantee 
the numerical errors will be small when the correctness of the reference solution is unknown.
A large self-consistency error is an indicator of numerical errors, but it might be that
small numerical errors initiated the solution to follow a different but still 
physically correct trajectory (to a good approximation). Therefore, attention should be
at the rate of initial deviation rather than the magnitude of the error over time. 

In \cite{Kadian2020}, it is argued that simulators need not be a perfect replica of
reality to be useful and are better judged by their sim-to-real predictivity: if one method 
outperforms another in simulation, how likely is the trend to hold in reality? For this 
purpose, they introduce a sim-to-real correlation coefficient (SRCC), which is the Pearson 
correlation coefficient over a set of performance pairs of reinforcement learning agents 
that are evaluated in simulation and reality. The sim-to-real gap vanishes as SRCC approaches 1.
Alternatively, the distance between predicted 
and real state-action transitions are measured \cite{allevato2020}.

In the field of machine learning, it is common to use either 
domain adaptation or domain randomization to reduce the effects of having a reality gap.
In \emph{domain adaptation}, a model learns about features invariant to the shift 
between training (simulation) and test (reality) distributions and uses this to generalize better under a domain shift.
\emph{Domain randomization} means that various attributes of 
the training domain are randomized to make the model more robust and adaptable to unseen and 
changing environmental conditions \cite{muratore2022}. With the wrong type or amount of randomization,
the model becomes overly conservative, or the problem becomes too hard.
In \cite{Tobin2017}, the reality gap from simulators with 
low-fidelity rendering was bridged by randomizing scene properties such as lighting, textures, and
camera placement. With the same approach, significant calibration errors in the dynamics model were mitigated
in \cite{Peng2018}. It has been suggested that domain randomization can effectively reduce 
simulation bias from numerical errors or unmodeled physics
when using a simulator of low fidelity.

\subsection{Measures}
In the present paper we consider scalar signals $f(t):[0,T]\to\mathbb{R}$, 
position trajectories $\bm{x}(t):[0,T]\to\mathbb{R}^{3}$, and discrete or 
time-integrated scalar quantities, $q$, such as loaded mass or total work. 
In time discrete form we represent the signals as $f_{0:T} = [f_{0},f_{1},...,f_{N}]$
and $\bm{x}_{0:T} = [\bm{x}_0,\bm{x}_1,...,\bm{x}_N]$, with 
the number of discrete timesteps $N = T/\Delta t$.
For each scalar signal $f_n$, with real reference $\hat{f}_n$, 
the instantaneous error at discrete time indexed $n$ is denoted $\varepsilon^f_n = f_n - \hat{f}_n$,
and we compute the normalized mean absolute error (MAE) by
\begin{equation}
  \mathcal{E}_f = \frac{1}{N}  \sum_{n=0}^{N}{\frac{\left|\varepsilon^f_n\right|}{f_\mathrm{norm}}},
\end{equation}
where $f_\mathrm{norm}$ is a normalizing reference value, which we take to be the maximum absolute
value if nothing else is mentioned.  For trajectories, a natural choice is the normalized mean Euclidean error (MEE)
\begin{equation}
  \mathcal{E}^{\sc{MEE}}_{\bm{x}} = \frac{1}{N}  \sum_{n=0}^{N}{\frac{\norm{\varepsilon^{\bm{x}}_n}_2}{L_\mathrm{norm}}},
\end{equation}
with Euclidean norm of the momentaneous trajectory error $\varepsilon^{\bm{x}}_n = \bm{x}_n - \hat{\bm{x}}_n$ 
and a normalizing length $L_\mathrm{norm}$.  However, if two trajectories trace approximately the same 
path with a slight delay or speed difference, this is picked up by the momentaneous error and accumulated along the reminding part of the trajectory.  

The dynamic time warping (DTW) distance \cite{Berndt1994} is a similarity measure useful for comparing trajectory time series. The cumulative error of a time shift or speed difference is much less than for the standard Euclidian norm, giving a more nuanced similarity measure. 
Assume two time discrete trajectories $\bm{x} = [\bm{x}_0,\bm{x}_1,...,\bm{x}_N]$ and $\bm{y} = [\bm{y}_0,\bm{y}_1,...,\bm{y}_N]$,
and a warping curve $\phi(n) = (\phi_{\bm{x}}(n),\phi_{\bm{y}}(n))$ with warping functions $\phi_x$ and 
$\phi_y$ that monotonically remaps the time series, i.e., $\phi_{\bm{x}}(n+1) \geq \phi_{\bm{x}}(n)$.  
The optimal warping curve picks the deformation of the time axis, which brings the two time series
as close as possible to each other, measured by
\begin{equation*}
    d_\phi (\bm{x},\bm{y}) = \sum_{n=0}^N \norm{\phi_{\bm{x}}(n)-\phi_{\bm{y}}(n)}_2.
\end{equation*}
We compute the normalized DTW distance error as
\begin{equation}
  \mathcal{E}^{\sc{DTW}}_{\bm{x}} = \frac{d_\phi(\bm{x},\hat{\bm{x}})}{N L_\mathrm{norm}}
\end{equation}
using the Python implementation \texttt
{similaritymeasures} from \cite{Jekel2019}. 

\section{Experiments}
\label{sec:experiment}

The experimental data in this study comes from a field test conducted by Komatsu Ltd
																																										
using a manually operated wheel loader equipped with additional sensing capabilities.
The vehicle, test environment, and the procedure for data collection and obtained 
measurements are described in this section.

\subsection{Wheel loader}
\label{sec:wheel-loader}
The vehicle was a Komatsu WA320-7, which is a medium-sized wheel loader commonly used 
in quarry mines and construction sites. It has an operating weight of 15.175 tonnes and is powered by a diesel engine with 
an output power capacity of 127 kW. A hydrostatic transmission driveline provides four-wheel drive via a 
fixed ratio gearbox and a differential system. The wheels, 1.39 m in 
diameter, are spaced at a track width of 2.05 m and a wheelbase of 3.03 m. 
A hydraulic-powered articulated steering joint separates the rear and front unit.
The flat-bottomed bucket has a loading capacity of 3.0 m$^3$ and is 2.685 m wide, 
																											 
and was equipped with a bolt-on cutting edge. The bucket is mounted on the front unit's parallel Z-bar linkage 
mechanism, which is hydraulically actuated with two (parallel) boom cylinders and one bucket 
cylinder for lifting and tilting. The test vehicle
was equipped with several sensing capabilities, including pressures in the lift and tilt hydraulic cylinders
and the geometric configuration of the bucket linkage.
The vehicle position and velocity
relative to the ground and walls confining the gravel pile were also tracked.
The vehicle control system for engine, transmission, and hydraulics balances the torque and fuel usage in the 
different work phases. The operator mainly controls the accel and brake pedals, shift range, and lift and tilt lever.

\subsection{Test environment}
The wheel loader was manually operated on the test site, which included a flat, rigid ground
and piles of gravel confined with vertical walls at the sides and the rear. 
The gravel consisted of sedimentary rock crushed and sifted to particle size 
around 30-40 mm mixed with a small amount of fine particles and moisture. The bulk 
density was measured to 1727 kg/m$^3$. An image of the wheel loader digging into the pile is shown in Fig.~\ref{fig:fieldtest} and in Supplementary Video 1.

\begin{figure}
  \centering
  \includegraphics[width=0.7\linewidth]{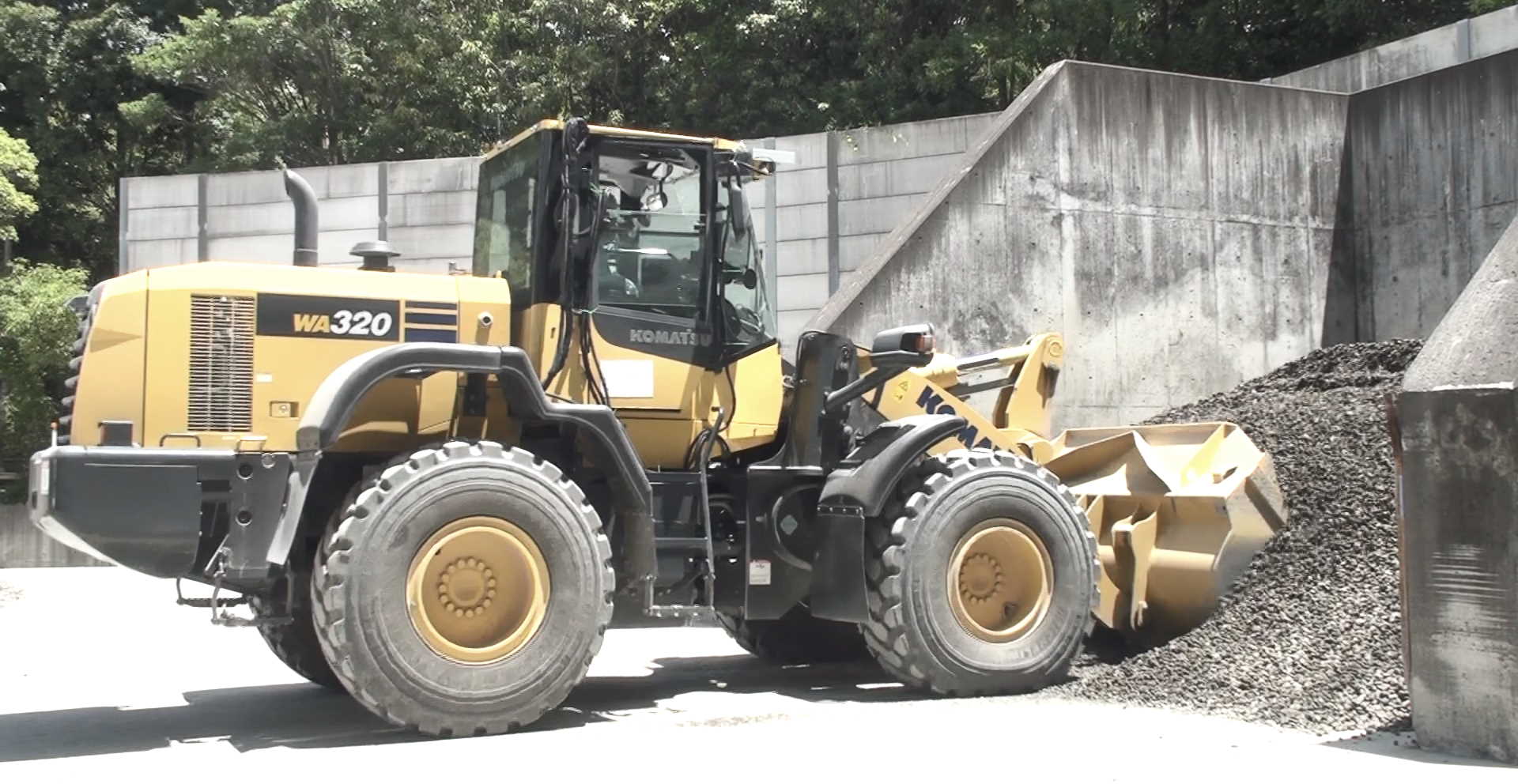}
  \caption{Photo from the field test.}
  \label{fig:fieldtest}
\end{figure}

\subsection{Measurements}
\begin{figure}[h]
  \centering
  \includegraphics[width=0.7\linewidth]{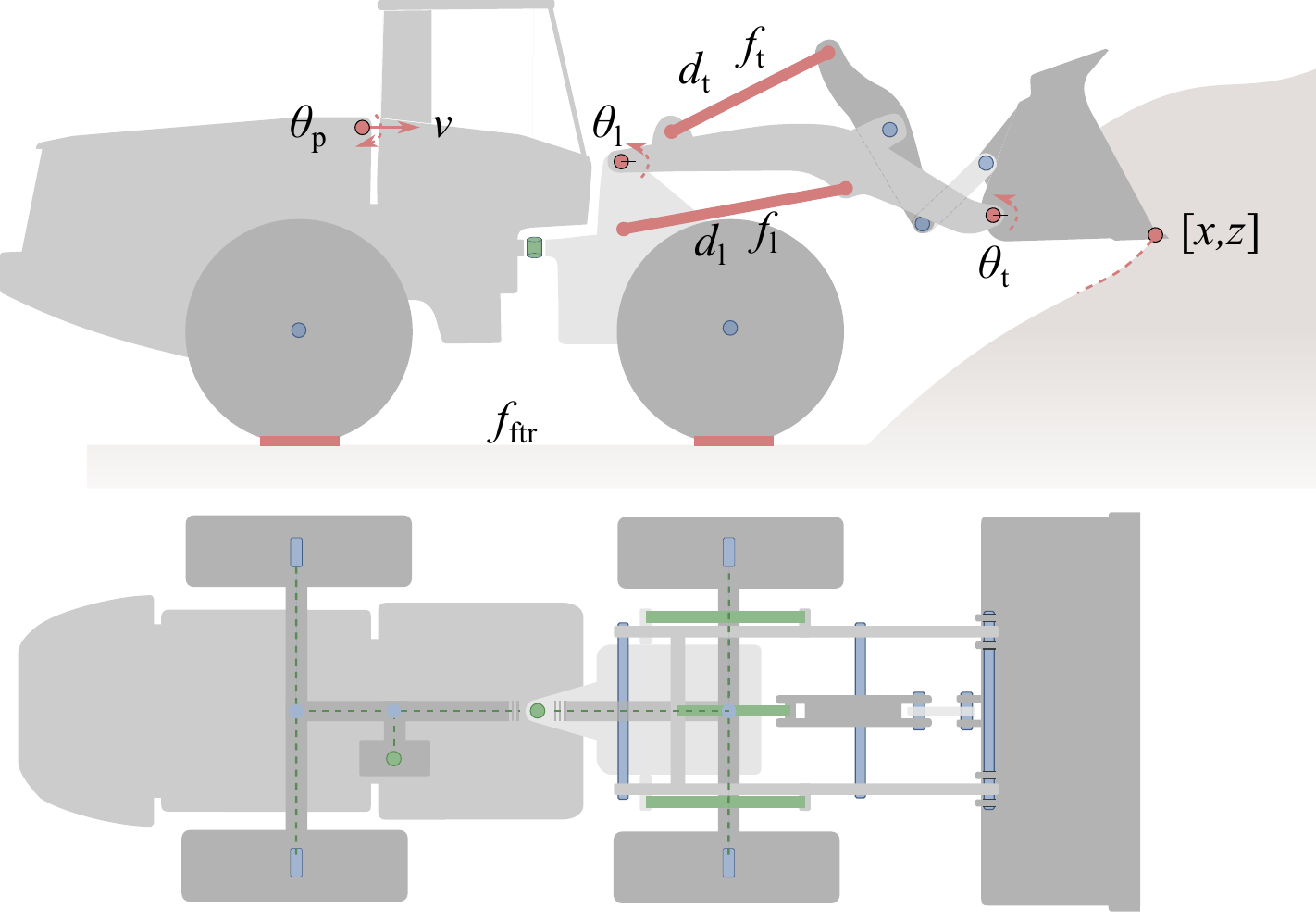}\\
  \caption{Illustration of the wheel loader from the side and the top with the quantities measured during the field test marked in red, joints in blue, and actuators in green.}
  \label{fig:fieldtest_measurement}
\end{figure}

Different loading operations were performed manually while collecting 
																								
measurement data with a sampling frequency of 100 Hz. This study focuses on the quantities listed in Table \ref{table:time_series} and illustrated
in Fig.~\ref{fig:fieldtest_measurement}.  Discrete measurements were also made. The shape of 
the pile surface was captured with a 2D laser scanner before each recorded loading. 
The loaded mass in the bucket was estimated after each loading using the
verified built-in functionality. The momentaneous power consumption is computed as the sum
of the tractive power plus the rate of work exerted by the lift and tilt cylinders,
$P \equiv P_\mathrm{tr} + P_\mathrm{l} + P_\mathrm{t} = f_\mathrm{tr}v + f_\mathrm{l}\dot{d}_\mathrm{l} + f_\mathrm{t}\dot{d}_\mathrm{t}$.
The reported net mechanical work is the time integral of this.
Note that the exerted work does not include the mechanical losses in the engine,
transmission, or hydraulics.

\begin{table}[ht] 
  \caption{Time series measurements.}
  \centering
  \begin{tabular}{|l|c|c|l|}
  \hline
  Variable          & 	Symbol  &  Unit  & Comment \\ 
  \hline
  lift force & $f_\mathrm{l}$     &  arb.u.   &  calculated from cylinder pressures \\
  tilt force & $f_\mathrm{t}$     &  arb.u.   &  calculated from cylinder pressures \\
  tractive force    &   $f_\mathrm{tr}$  &  arb.u. & calculated from hydraulic motor pressure and capacity \\
  drive speed    &   $v$  &  km/h & calculated from wheel revolution \\ 
  lift extension & $d_\mathrm{l}$     &  arb.u.   &  calculated from the piston position \\
  tilt extension & $d_\mathrm{t}$     &  arb.u.   &  calculated from the piston position \\
  boom angle    &   $\theta_\mathrm{l}$  &  degrees & angular potentiometer relative to chassis pitch \\ 
  bucket angle    &   $\theta_\mathrm{t}$  &  degrees & angular potentiometer relative to chassis pitch \\ 
  chassis pitch  &   $\theta_\mathrm{p}$  &  degrees & calculated from the tire strains relative to ground \\  
  vehicle position  &    &  m & laser scanned distance  to the vertical wall behind the pile \\    
														  
  bucket position &   $[x,z]$  &  m & calculated from  vehicle, boom, and bucket pose \\    \hline
\end{tabular}
\label{table:time_series}
\end{table}   

Three loading operations were selected for comparison with simulations.
These are listed in Table \ref{table:field_test}. In the test named FB35, 
the bucket was lowered horizontally to the ground and filled by 
driving it deep into the pile. After tilting the bucket $20^\circ$, 
the wheel loader was reversed. After breakout, the bucket was finally tilted to the end. 
During the penetration phase, a slight increase in the boom lift was applied (starting at $t = 3.5$ s) 
to avoid wheel slip when maximum traction is required. After loading was completed, the weight of the soil in the bucket was estimated to $3.46$ tonnes and the exerted mechanical work was $209$ kJ. 
In the HD27 test, the boom was raised during the middle of the penetration phase 
(between $t = 1.5$ and $t = 2.5$ s) and tilted to the end during breakout. The 
loaded mass was $2.70$ tonnes, and the net work was $127$ kJ.
The RD21 test is characterized by a shallow bucket penetration at low speed with the
bucket tip raised 0.25 m above the ground. A step-wise tilting of the bucket was applied during
bucket filling. The operation resulted in $2.10$ tonnes of 
loaded mass and a mechanical work of $112$ kJ. 
The measurements from the three loadings are presented in Fig.~\ref{fig:trajectories}
and \ref{fig:force_velocity} together with simulated measurements.
Note that these operations were the result of trying different trajectories with no ambition of achieving the rated load.

\begin{table}[ht] 
  \caption{Loaded mass and estimated work in the three field experiments.}
  \centering
  \begin{tabular}{|l|c|c|}
  \hline
  Test     & 	Mass [tonne]  & Work [kJ] \\ 
  \hline
  FB35     &   $3.46$   &  $209$ \\
  HD27     &   $2.70$   &  $127$  \\
											   
  RD21      &   $2.10$   &  $112$ \\
  \hline
\end{tabular}
\label{table:field_test}
\end{table}   

\section{Simulator}
\label{sec:simulator}
\begin{figure}
  \centering
  \begin{subfigure}[b]{0.45\textwidth}
      \centering
      \includegraphics[width=1.0\linewidth,trim={0mm 0mm 0mm 0mm},clip]{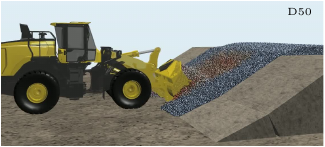}
      \includegraphics[width=1.0\linewidth,trim={0mm 0mm 0mm 0mm},clip]{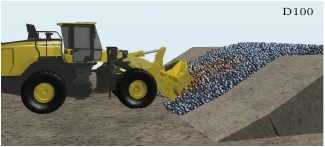}
      \includegraphics[width=1.0\linewidth,trim={0mm 0mm 0mm 0mm},clip]{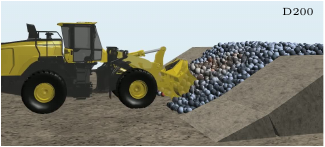}
      \includegraphics[width=1.0\linewidth,trim={0mm 0mm 0mm 0mm},clip]{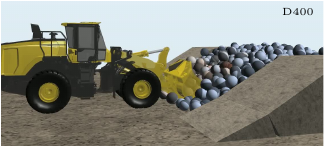}
  \end{subfigure}
  \begin{subfigure}[b]{0.45\textwidth}
    \centering
    \includegraphics[width=1.0\linewidth,trim={0mm 0mm 0mm 0mm},clip]{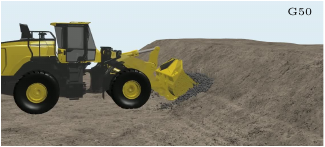}
    \includegraphics[width=1.0\linewidth,trim={0mm 0mm 0mm 0mm},clip]{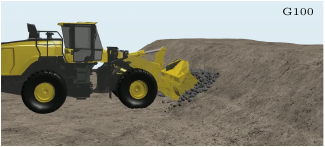}
    \includegraphics[width=1.0\linewidth,trim={0mm 0mm 0mm 0mm},clip]{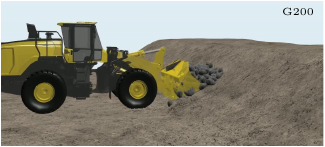}
    \includegraphics[width=1.0\linewidth,trim={0mm 0mm 0mm 0mm},clip]{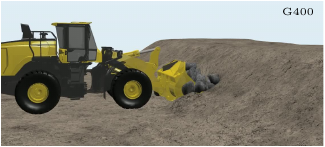}
\end{subfigure}
\caption{Images from the eight simulators of different levels of fidelity. In the type D-simulators (left column), 
     the gravel pile is fully resolved in particles with characteristic diameters of 50, 100, 200, and 400 mm (top to bottom).
     In the type G-simulators (right column), a multiscale technique is applied with different grid sizes 50, 100, 200, and 400 mm (top to bottom).}
     \label{fig:simulator_fidelity}
\end{figure}

Simulators were created for the loading scenarios in Sec.~\ref{sec:experiment}.
The simulators include a wheel loader, a rigid flat ground, and a pile of soil. 
The vehicle is modeled as a 3D rigid multibody system with frictional contacts
and driveline dynamics.
For the soil, two different types of models are used, type D and G,
each with four different spatio-temporal resolutions. 
The key settings for the eight different levels of simulator fidelity are 
listed in Table \ref{table:simulators}, with characteristic particle diameter $D$, 
timestep $\Delta t$, the number of particles $N_\mathrm{p}$, and the number 
of solver iterations $N_\mathrm{it}$ (explained in Sec.~\ref{sec:particle_terrain}). 
Screen captures from the eight simulators are shown in Fig.~\ref{fig:simulator_fidelity}, and Supplementary Video 2 shows the evolution.
In the type-D simulators, the entire pile is modeled in terms of particles using 
DEM with time implicit integration for strong coupling with 
the vehicle dynamics through the particle-bucket contact forces.
These simulations are computationally intense, 
especially when the soil is finely resolved into many small particles.
In the type-G simulators, a reduced multiscale method is used where 
only a small fraction of the soil, the active zone inside and in front of the bucket, is resolved 
in terms of particles. The macroscopic dynamics of the particle system is
approximated by a rigid aggregate body that is coupled back to the vehicle dynamics.
The type-G simulators are computationally much more efficient, running in real-time or faster 
when the grid size is set large enough but presumably associated with a larger model error.
The simulations were performed using the physics engine AGX Dynamics \cite{AGX20} 
with the methods described in \cite{servin:2014:esn} and \cite{Servin2021}.
Details are described in the following subsections.

\begin{table}[ht] 
  \caption{Key settings for the eight different levels of simulator fidelity shown in Fig.~\ref{fig:simulator_fidelity}.}
  \centering
  \begin{tabular}{|l|r|r|r|r|}
  \hline
  Fidelity level & $D$ [mm]  &  $\Delta t$ [ms]  & $N_\mathrm{p} \quad$ & $N_\mathrm{it}$\\ 
  \hline
  D50           & 50              &  2              &  $450,000$     &  200 \\
  D100          & 100             &  5              &  $55,000$     &  100 \\
  D200          & 200             &  10             &  $6,100$     &  500 \\
  D400          & 400             &  20             &  $800$     &  15 \\
  \hline
  G50           & 50              &  5              &  ${\scriptstyle\lesssim}\, 28,000$     &  50 \\
  G100          & 100             &  10             &  ${\scriptstyle\lesssim}\, 3,500$     &  25 \\
  G200          & 200             &  20             &  ${\scriptstyle\lesssim}\, 480$     &  15 \\
  G400          & 400             &  50             &  ${\scriptstyle\lesssim}\, 55$     &  10 \\
  \hline
\end{tabular}
\label{table:simulators}
\end{table}

\subsection{Multibody dynamics framework}
\label{sec:MBD}
We use non-smooth contacting multibody dynamics in descriptor form for modeling the vehicle and the soil, introduced in \cite{Lacoursiere2007} and supported by AGX Dynamics \cite{AGX20}. 
Specifically, we use a maximal coordinate representation in terms of rigid bodies and various types of kinematic constraints for joints, motors, and frictional contacts and impacts. The governing equations are
\begin{align}
  \label{eq:momentum}
  \bm{M} \dot{\bm{v}}  = \bm{f} + \bm{G}_\mathrm{j}^T \bm{\lambda}_\mathrm{j} + \bm{G}_\mathrm{c}^T \bm{\lambda}_\mathrm{c},\\
  \label{eq:kinematic_constraint}
  \varepsilon_{\mathrm{j}} \bm{\lambda}_{\mathrm{j}} + 
        \eta_{\mathrm{j}} \bm{g}_{\mathrm{j}} + 
        \tau_{\mathrm{j}} \bm{G}_{\mathrm{j}} \bm{v} = \bm{u}_{\mathrm{j}}, \\   \label{eq:limits}
    \bm{\lambda}_\mathrm{min} \leq \bm{\lambda}_\mathrm{j} \leq \bm{\lambda}_\mathrm{max},\\
  \label{eq:contact_constraint}\mathrm{contact\_law(}\bm{g}_{\mathrm{c}},\bm{v}_{\mathrm{c}},\bm{\lambda}_{\mathrm{c}} \mathrm{)},
\end{align}
with system mass matrix $\bm{M} \in \mathbb{R}^{6N_\mathrm{b}\times6N_\mathrm{b}}$, external force $\bm{f}\in\mathbb{R}^{6N_\mathrm{b}}$, and velocity $\bm{v}\in\mathbb{R}^{6N_\mathrm{b}}$ that is the time derivative of the world frame maximal coordinates $\bm{x}\in\mathbb{R}^{7N_\mathrm{b}}$ (using quaternions for the orientation). The constraint forces in the Newton-Euler equation of motion (\ref{eq:momentum}), with Lagrange multiplier $\bm{\lambda}$ and Jacobian $\bm{G}$ are divided
joints and motors, labeled with $\mathrm{j}$, and contacts, labeled with $\mathrm{c}$.
Eq.~\eqref{eq:kinematic_constraint} is a generic constraint equation.  An ideal joint can be represented with $\varepsilon_{\mathrm{j}} = \tau_{\mathrm{j}} = \bm{u}_{\mathrm{j}} = 0$, in which case Eq.~\eqref{eq:kinematic_constraint} express a holonomic constraint, $\bm{g}_{\mathrm{j}}(\bm{x}) = 0$.  
A non-ideal joint is modeled using finite compliance $\varepsilon_{\mathrm{j}}$ and viscous damping rate $ \tau_{\mathrm{j}}$.
A linear or angular motor may be represented by a velocity constraint $ \bm{G}_{\mathrm{j}} \bm{v} = \bm{u}_{\mathrm{j}}(t)$ with target speed $\bm{u}_{\mathrm{j}}(t)$, which follows by  $\varepsilon_{\mathrm{j}} = \eta_{\mathrm{j}} = 0$ and $ \tau_{\mathrm{j}} = 1$. Range limits on the motor constraint forces may be imposed by Eq.~\eqref{eq:limits}.
With $N_\mathrm{j}$ constrained and actuated degrees of freedom we have $\bm{\lambda}_{\mathrm{j}}\in\mathbb{R}^{N_\mathrm{j}}$ and $\bm{G}_\mathrm{j}\in\mathbb{R}^{N_\mathrm{j}\times6N_\mathrm{b}}$.

Contact laws are imposed as inequality and complementarity conditions on the contact multiplier $\bm{\lambda}_{\mathrm{c}}\in\mathbb{R}^{3N_\mathrm{c}}$
and relative contact velocity.
Each contact multiplier 
is split $\bm{\lambda}_{\mathrm{c}} = [\lambda_{\mathrm{n}};\bm{\lambda}_{\mathrm{t}}]$
in the normal and tangential components that must obey the Coulomb law, 
$|\bm{\lambda}_{\mathrm{t}}| \leq \mu_{\mathrm{t}} \lambda_{\mathrm{n}}$, and complementarity conditions for non-penetration $0\leq \varepsilon^{-1}_\mathrm{c}g_\mathrm{n} + \gamma_\mathrm{c} \bm{G}_\mathrm{n}\bm{v} \perp \lambda_\mathrm{n}\geq 0$ and no-slip $\norm{\bm{G}_\mathrm{t} \bm{v}}(\mu_{\mathrm{t}} \lambda_{\mathrm{n}} - \norm{\bm{\lambda}_{\mathrm{t}}}) $, and maximumum dissipation $\norm{\bm{G}_\mathrm{t} \bm{v}}\norm{\bm{\lambda}_{\mathrm{t}}} = - (\bm{G}_\mathrm{t} \bm{v})^T\bm{\lambda}_{\mathrm{t}}$. Here, $g_\mathrm{n}$ is a contact gap function, with normal Jacobian $\bm{G}_\mathrm{n}=\frac{\partial g_\mathrm{n}}{\partial \bm{x}}$, contact compliance $\varepsilon_\mathrm{n}$ and damping $\gamma_\mathrm{n}$, and the tangential Jacobian $\bm{G}_\mathrm{t}$ is such that the contacting bodies relative velocity in the contact tangent space is given by $\bm{G}_\mathrm{t} \bm{v}$. 
The set of active contacts in Eq.~\eqref{eq:contact_constraint}, with normal gaps overlaps stored in $\bm{g}_\mathrm{c}$ and contact velocities $\bm{v}_\mathrm{c} \equiv \bm{G}_\mathrm{c} \bm{v} = [\bm{G}^T_\mathrm{n},\bm{G}^T_\mathrm{t}]^T \bm{v}$, are re-computed at every simulation timestep using a collision detection algorithm.
High-velocity impacts are modeled using the Newton impact law while preserving all other kinematic constraints. 
For the DEM bodies (soil particles), the contact model is mapped to Hertz model and a rolling resistance constraint is included to capture the effect of the real particles having a non-spherical shape while the simulated particles are spherical. The details about the mapping and parametrization of the contact model and the Jacobians are found in \cite{servin:2014:esn,wiberg:2021:dem,Servin2021}.

The dynamics system is time-integrated using the SPOOK stepper \cite{Lacoursiere2007}, which is a first-order accurate discrete variational integrator developed particularly for fixed timestep real-time simulation of multibody systems with non-ideal constraints and non-smooth dynamics. The time-discrete equations, forming a mixed complementarity problem (MCP), are solved using the direct-iterative split solver in AGX \cite{AGX20}. 
A block-sparse LDLT solver with pivoting \cite{lacoursiere2010} is used as direct solver
for the vehicle system and its external contacts, with linearization of the Coulomb friction model. The DEM equations are solved using a projected Gauss-Seidel (PGS) algorithm, which is accelerated using domain decomposition for parallel processing and warm-starting \cite{wang:2016:wsp}. The latter solves the contact problem without linearization of the Coulomb law.

\subsection{Wheel loader model}
\label{sec:wheel-loader-model}
The vehicle simulation model matches the key geometric dimensions and mass distribution of 
the Komatsu WA320-7 wheel loader introduced in Sec.~\ref{sec:wheel-loader}, and shown in Fig.~\ref{fig:fieldtest_measurement}. It is composed
of ten rigid bodies, ten hinge joints, and three prismatic joints. The mass and geometry properties were determined from a 3D model and geometric information provided by the manufacturer. The lift and tilt
hydraulic cylinders are modeled as independent linear motors, introduced as velocity constraints through Eq.~(\ref{eq:kinematic_constraint}).
The cylinders are controlled by assigning a momentaneous target speed and a maximum force that is derived 
from the manufacturer's specifications. The resulting actuator speed and applied force depend on the dynamic 
state and are computed by the multibody dynamics solver. 
The vehicle model is equipped with a minimalistic driveline model. The engine is
modeled as a hinge motor constraint with a torque limit that depends on the 
rotational speed. A set target drive speed is translated into a target motor speed. 
The rotational motion is transmitted to the wheels via a main drive shaft and differentials.
For the tire-ground contacts, the linear elastic modulus was set to $1.0$ MPa and the surface friction coefficient to $2.0$. The elasticity value was found to best match the experimentally observed tire deflection during bucket filling. The friction coefficient was the lowest value found that did not cause tire slippage during bucket filling in the calibration experiments. Having a friction coefficient larger than unity is not uncommon for tires on rough surfaces. Internal friction in joints and hydraulic cylinders is assumed to be negligible compared to the dig forces and was not modeled.

\subsection{Particle terrain model}
\label{sec:particle_terrain}
In the type-D simulators, the entire pile is resolved into particles that are 
simulated using the nonsmooth DEM as described in section \ref{sec:MBD} and more detailed in \cite{servin:2014:esn} and \cite{wiberg:2021:dem}.
The piles are created by emitting particles into a 6 m wide container with a front surface
shaped as in the field tests. The particles are given a spherical shape, a specific mass density $2590$ 
																							  
kg/m$^3$, friction coefficient $0.3$, rolling resistance coefficient $0.02$, and zero restitution coefficient. This matches the field test bulk mass density $1727$ kg/m$^3$ and the $32^{\circ}$ 
angle of repose, the best among the pre-calibrated soils in \cite{servin:2014:esn}.
For each field test, three piles were created with different particle size, $D = 50, 100, 200$, and $400$ mm.
To avoid the formation of regular packings, the particle diameters are slightly perturbed into
a uniform size distribution in a small size span of $D \pm 0.1D$. With nonsmooth DEM, the computational time 
is proportional to $N_\mathrm{p}N_\mathrm{it}/\Delta t \propto D^{3.5}$. The strong
dependency on the spatial resolution follows from the empirical rules 
$N_\mathrm{it} \gtrsim 0.1 (L/D)/\epsilon$ and $\Delta t \lesssim \sqrt{2\epsilon D/g}$ 
for obtaining an error tolerance $\epsilon$ when simulating granular systems with characteristic size $L$ and 
gravity acceleration $g$, using the SPOOK stepper and the projected Gauss-Seidel (PGS) solver \cite{servin:2014:esn}.

\subsection{Multiscale terrain model}
\begin{figure}
  \centering
  \includegraphics[width=0.85\linewidth,trim={0mm 0mm 0mm 0mm},clip]{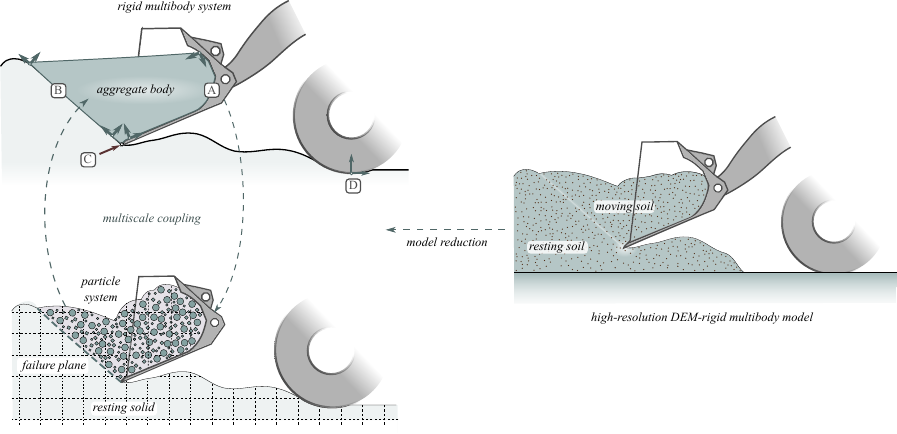}
  \caption{Illustration of the multiscale terrain model (left) adapted from \cite{Servin2021}. It can be regarded as a heavily reduced version of the full DEM model (right). The region of active soil movement is predicted and substituted with a rigid aggregate body that couples with the vehicle dynamics (upper left). The mass flow inside the active region is co-simulated using a DEM model (lower left). }
     \label{fig:multiscale}
\end{figure}

The type-G simulators use the multiscale model for deformable terrain described in 
\cite{Servin2021} and illustrated in Fig.~\ref{fig:multiscale}. 
This model can be understood as a heavily reduced approximation of the full DEM model. From the perspective of the vehicle, the region of active soil is substituted with a single rigid body that has contact support with the surrounding terrain of resting soil. 
The coupled dynamics of the vehicle and the soil aggregate body are modeled using Eqs.~(\ref{eq:momentum})-(\ref{eq:contact_constraint}) and solved with high accuracy using the direct solver.

The terrain deformations inside the active zone are treated by co-simulated soil dynamics models with the vehicle represented as kinematic bodies. The procedure is as follows. 
The surface of the terrain is represented by a heightmap. Its initial shape is reconstructed from the scanned piles in the field test. In solid phase, the soil
is represented using a regular grid of voxels with variable states of mass occupancy and 
compaction. It is assigned a set of bulk mechanical parameters for its physical
behavior in a nominal bank state. 
When a bucket comes in contact with the terrain surface, the zone of active soil 
movement is predicted. It is comprised of a shear failure plane stretching from the 
bucket's cutting edge to the soil surface, enclosing a soil wedge. The failure angle depends on the soil's internal 
friction and on the orientation of the bucket's cutting plane.
Inside the active zone, the soil is represented primarily by particles that may grow and 
shrink in size and numbers as the active zone progresses into or out of the terrain along with the moving bucket.
							 
The particle dynamics is modelled using DEM with specific mass density and contact 
parameters that ensure a bulk mechanical behavior consistent with the set bulk parameters. 
The particles evolve in a co-simulation where the bucket exists as a kinematic moving body, controlled by the vehicle multibody simulation. The reaction force on the bucket from the soil is mediated through an aggregate body
																					   
that inherits the momentaneous shape, inertia, and momentum of the particles in the active zone.
				
The soil's internal friction is applied at the aggregate-terrain contact
interface (B in Fig.~\ref{fig:multiscale}) while distinct parameters may be set for the
aggregate-bucket interface (A in Fig.~\ref{fig:multiscale}).  

The role of the aggregate body can be viewed as a multibody dynamics generalization of the 
soil separation force described by the fundamental earthmoving equation \cite{mckyes:1985:sct}.
Besides capturing inertial effects, the aggregate body has a numeric filtering effect
that provides a stable reaction force and rate of soil displacement despite the large stresses and coarse spatial and temporal resolution.
For this reason, it is possible to simulate with a larger timestep and 
less PGS solver iterations than predicted by the relations in Sec.~\ref{sec:particle_terrain} for DEM.

The additional resistance for the bucket teeth or edge to penetrate dense soil
under stress is modeled by a penetration constraint (C in Fig.~\ref{fig:multiscale}) that hinders the motion of the bucket
in its cutting direction unless the penetration resistance force exceeds a critical value
which is a function of the bucket geometry and the soil-bucket surface friction coefficient.

The terrain model was assigned the bulk parameters that best match the observed
properties of the gravel at the field test site: mass density $1727$ kg/m$^3$, 
internal friction angle $32^\circ$, dilatancy angle $8^\circ$,  
and Young's modulus $4.6$ MPa. Dilatancy is the volume expansion in a granular media that is induced by shear deformation. It makes the material more resistant to shear failure by adding directly to the internal friction, i.e., summing to an effective internal friction of $40^\circ$ in the present case. The bucket cutting edge was assigned a maximum and minimum radius 
of $10$ mm and $2.5$ mm, respectively, and an equivalent tooth length of $10$ mm. Two parameters were
calibrated for the best match between the simulated and measured dig forces. 
The friction coefficient between the bucket and soil was set to $0.2$. The so-called 
aggregate stiffness multiplier was set to $0.01$. This increases the contact elasticity
at the aggregate-terrain interface by a factor of five relative to the set Young's modulus of the soil. 

\begin{figure}
  \centering
  \includegraphics[height=0.125\linewidth]{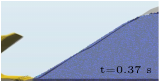}
  \hspace{-2mm}
  \includegraphics[height=0.125\linewidth]{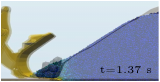}
  \hspace{-2mm}
  \includegraphics[height=0.125\linewidth]{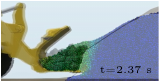}
  \hspace{-2mm}
  \includegraphics[height=0.125\linewidth]{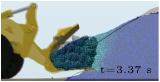}
  \\
  \includegraphics[height=0.125\linewidth]{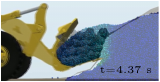}
  \hspace{-2mm}
  \includegraphics[height=0.125\linewidth]{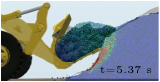}
  \hspace{-2mm}
  \includegraphics[height=0.125\linewidth]{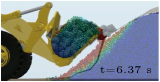}
  \hspace{-2mm}
  \includegraphics[height=0.125\linewidth]{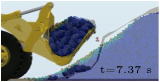}
  \\
  \includegraphics[height=0.125\linewidth]{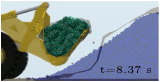}
  \hspace{-2mm}
  \includegraphics[height=0.125\linewidth]{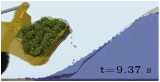}
  \hspace{-2mm}
  \includegraphics[height=0.125\linewidth]{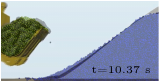}
  \hspace{-2mm}
  \includegraphics[height=0.125\linewidth]{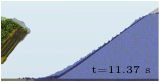}
  \caption{Simulation of the HD27 test with overlaid images from the D50 and G200 simulators at one-second intervals starting from time $0.37$ s.  The D50 particles are color-coded by speed, with blue for 0 and red for 1 m/s, while the G200 particles are gray. 
  The shape of the active zone and the distributions of mass in the bucket are in good agreement. }
     \label{fig:hd27-D50G200}
\end{figure}

In Fig.~\ref{fig:hd27-D50G200} and Supplementary Video 3, the evolution of the HD27 test using the G200 
simulator is compared with using a D50 simulator, the one with finest resolution.
The simulators differ in number of particles by a factor $10^3$ and in computational speed by $10^4$. 
Despite this, there is good agreement in the evolution of the G200 active zone and the mobilized D50 particles, and in the distributions of soil in the bucket after breakout.
The differences in soil model and spatio-temporal resolution produce a small offset in the
vehicle poses. This appears as a slight double vision in the images.

\section{Comparison}
\label{sec:comparison}

\subsection{Feedforward control}
\label{sec:feedforward}
To assess the reality gap of the simulators with different levels of fidelity, we repeated the loading cycles
from the field test using feedforward control of the forward drive, boom lift, and bucket tilt actuators using target speed time series as input. The target speed control signals that best replicated the motion of the operator-controlled vehicle were identified using the G200 simulator (as it runs in real-time).  This stage

also involved calibration of the wheel-ground friction coefficient, the bucket-terrain friction 
coefficient, and the aggregate stiffness multiplier to the values listed in the previous section. 
Next, the identified feedforward control signals (actuator target speed) were used as input when running
each simulator. 

\subsection{Processing of time series}
\label{sec:time series}
The comparison between the time series measurement from simulations and field tests
was made with the observation variables listed in Table \ref{table:time_series}. 
Force measurements are rescaled by dividing them by a force constant characteristic for the vehicle.
We exclude the phases of initializing the vehicle to target speed before reaching the pile and the phase after the bucket reaches the end. These phases are indicated in gray in Fig.\ref{fig:force_velocity}.

For the scalar signals, the MAE error was computed, while the DTW error was used for the bucket tip trajectories.
The errors for each variable and D and G simulator are found in Table \ref{table:sim-to-real_error_granular}
and \ref{table:sim-to-real_error_terrain}, respectively. 
These tables also include the relative errors in loaded mass $\mathcal{E}_M = (M - \hat{M})/\hat{M}$, work, $\mathcal{E}_W = (W - \hat{W})/\hat{W}$, and the mean error for each simulator.

\begin{table} 
  \caption{Errors of the type-D simulators.}
   \tiny
    \centering
    \begin{filecontents*}{fig/csv/D-table.csv}
name,RMA-traj,RMA-V,RMA-TR,RMA-L,RMA-T,load_mass,work,E-score
FB35-D50,0.0984290726953231,0.11951503267973856,0.09054086054117647,0.09478856205949525,0.09005927595411155,-0.19942196531791906,-0.22104865956333195,0.1305433469730137
FB35-D100,0.038962859228437785,0.1217986928104575,0.0876039,0.09126940599975221,0.09929890804723426,-0.15317919075144504,-0.1958954526366044,0.1125726299248473
FB35-D200,0.030417981961186984,0.12349607843137254,0.08383484310588234,0.10526514163296796,0.09620464397653464,-0.15317919075144504,-0.12146915202395875,0.10198100455476403
FB35-D400,0.10431925820607055,0.13221895424836602,0.08245510929411765,0.09780620901361475,0.13227397255553,-0.22254335260115607,-0.15885699108756418,0.1329248352866313
HD27-D50,0.02135554922763817,0.06039023297491039,0.13372297203225808,0.080630462606196,0.07612721554666167,-0.05185185185185189,0.14451308484284967,0.08122733844033797
HD27-D100,0.0213601243356781,0.05530421146953405,0.15528017812903225,0.07177338294082329,0.08918411380258753,0.01,0.1724440054465958,0.08182191764737978
HD27-D200,0.0322783392122853,0.06749955197132616,0.1444594413548387,0.11896736382199787,0.1570576880630336,-0.07037037037037051,0.04078550198781434,0.09020260811166664
HD27-D400,0.048559035108356366,0.05896146953405018,0.12536466341935484,0.11199778541349563,0.17976022198124478,0.01,-0.161791641700927,0.09912031779497658
RD21-D50,0.012528074647773047,0.04260069444444444,0.08126807168333333,0.07797139354731174,0.0908074627741182,-0.2571428571428572,-0.24617034486958836,0.11549841415848949
RD21-D100,0.024386760790202213,0.04635902777777778,0.08766865703333333,0.10239618462443606,0.1303768297658559,-0.31428571428571433,-0.24802531806674477,0.13621407033486632
RD21-D200,0.016434236157349975,0.04225717592592593,0.0709763076,0.10138110727829118,0.14969864882557612,-0.26666666666666666,-0.31447760463321395,0.1374131067267177
RD21-D400,0.08109005161147588,0.06290925925925926,0.12060800675000001,0.15993641384988647,0.19822033921430907,-0.30000000000000004,-0.6064709050298827,0.21846213938783046
\end{filecontents*}
\pgfplotstabletypeset[
      multicolumn names, 
      col sep=comma, 
      display columns/0/.style={
        column name=simulation, 
        string type,
        column type = {l}},  
      display columns/1/.style={
        column name=$\mathcal{E}_{\bm{x}}$,
        std, std=-2:1, precision=2, fixed zerofill,
        column type = {r}},
      display columns/2/.style={
        column name=$\mathcal{E}_v$,
        std, std=-2:1, precision=2, fixed zerofill,
        column type = {r}},
      display columns/3/.style={
        column name=$\mathcal{E}_{\text{tr}}$,
        std, std=-2:1, precision=2, fixed zerofill,
        column type = {r}},
      display columns/4/.style={
        column name=$\mathcal{E}_l$,
        std, std=-2:1, precision=2, fixed zerofill,
        column type = {r}},
      display columns/5/.style={
        column name=$\mathcal{E}_t$,
        std, std=-2:1, precision=2, fixed zerofill,
        column type = {r}},
      display columns/6/.style={
          column name=$\mathcal{E}_M$,
          std, std=-2:1, precision=2, fixed zerofill,
          column type = {r}},
      display columns/7/.style={
          column name=$\mathcal{E}_W$,
          std, std=-2:1, precision=2, fixed zerofill,
          column type = {r}},
      display columns/8/.style={
        column name=mean, 
        std, std=-2:1, precision=2, fixed zerofill,
        column type = {r}},
      every head row/.style={
      before row={\toprule}, 
      after row={\midrule},
      column type = {r},
            },
      every row no 4/.style={before row=\hline},
      every row no 8/.style={before row=\hline},
      every last row/.style={after row=\bottomrule}, 
        std, std=-2:1, precision=2, fixed zerofill,
    ]{fig/csv/D-table.csv} 
\label{table:sim-to-real_error_granular}
\end{table}

\begin{table}
  \caption{Errors of the type-G simulators.}
   \tiny
    \centering
    \begin{filecontents*}{fig/csv/G-table.csv}
name,RMA-traj,RMA-V,RMA-TR,RMA-L,RMA-T,load_mass,work,E-score
FB35-G50,0.10699366820521247,0.15669150326797385,0.07778662552941176,0.08560283928660092,0.13265650002704077,-0.05202312138728328,0.0,0.08767990742773625
FB35-G100,0.04043118467455896,0.14465947712418298,0.07024259188235295,0.08746483116171239,0.14150370466077888,0.0,0.012809735763695616,0.07101593218104024
FB35-G200,0.06317120206355442,0.11956535947712416,0.09248907889411764,0.09250077918629336,0.14845120796254432,0.0,-0.06649952065467214,0.08365247452124437
FB35-G400,0.192901540191246,0.24227450980392154,0.09260705629411764,0.12200991414007592,0.15399549037135057,-0.12427745664739889,-0.1166175405777291,0.14924050114654852
HD27-G50,0.0692881976831618,0.077790770609319,0.1888542342903226,0.15559023185734208,0.11573295196934004,0.1888888888888888,0.3288172665134629,0.16070893454454815
HD27-G100,0.04983482775013134,0.07474327956989246,0.12612839729032257,0.10998516408543665,0.10404540463157487,0.14444444444444432,0.13173879105622233,0.1058457584040035
HD27-G200,0.031110596895203006,0.06993996415770608,0.11707769651612901,0.08992744210660007,0.10200279703497421,0.13333333333333328,0.0,0.07786314538954257
HD27-G400,0.11451548475710817,0.07208109318996415,0.09635041572580645,0.08543513656509444,0.11037802970927076,0.040740740740740695,-0.15296750963401795,0.09606691576028607
RD21-G50,0.04804311606248851,0.04505625,0.15760849415,0.06586031907714057,0.0714329112925322,0.09999999999999998,0.15690007482941223,0.09212873791593905
RD21-G100,0.05895020804371135,0.04627337962962962,0.13807587921666667,0.05226327906510219,0.07718033610387609,0.10952380952380951,-0.0319564745569417,0.07346048087710529
RD21-G200,0.10023972806414042,0.04349907407407407,0.10323973708333334,0.05612970662439717,0.07049128003833481,0.20476190476190462,-0.20066992601405145,0.11129019380860512
RD21-G400,0.2145978328882501,0.04247453703703703,0.16273347025,0.09557448287902533,0.07719323307963476,0.3095238095238095,-0.3083002450868257,0.17291394439208319
\end{filecontents*}
\pgfplotstabletypeset[
      multicolumn names, 
      col sep=comma, 
      display columns/0/.style={
        column name=simulation, 
        string type,
        column type = {l}},  
      display columns/1/.style={
        column name=$\mathcal{E}_{\bm{x}}$,
        std, std=-2:1, precision=2, fixed zerofill,
        column type = {r}},
      display columns/2/.style={
        column name=$\mathcal{E}_v$,
        std, std=-2:1, precision=2, fixed zerofill,
        column type = {r}},
      display columns/3/.style={
        column name=$\mathcal{E}_{\text{tr}}$,
        std, std=-2:1, precision=2, fixed zerofill,
        column type = {r}},
      display columns/4/.style={
        column name=$\mathcal{E}_l$,
        std, std=-2:1, precision=2, fixed zerofill,
        column type = {r}},
      display columns/5/.style={
          column name=$\mathcal{E}_t$,
          std, std=-2:1, precision=2, fixed zerofill,
          column type = {r}},
      display columns/6/.style={
          column name=$\mathcal{E}_M$,
          std, std=-2:1, precision=2, fixed zerofill,
          column type = {r}},
      display columns/7/.style={
          column name=$\mathcal{E}_W$,
          std, std=-2:1, precision=2, fixed zerofill,
          column type = {r}},
      display columns/8/.style={
        column name=mean,
        std, std=-2:1, precision=2, fixed zerofill,
        column type = {r}},
      every head row/.style={
      before row={\toprule}, 
      after row={\midrule},
      column type = {r},
            },
      every row no 4/.style={before row=\hline},
      every row no 8/.style={before row=\hline},
      every last row/.style={after row=\bottomrule}, 
        std, std=-2:1, precision=2, fixed zerofill,
    ]{fig/csv/G-table.csv} 
\label{table:sim-to-real_error_terrain}
\end{table}

\subsection{Bucket tip trajectories}
\label{sec:trajectories}
\begin{figure}[!htb]
  \centering
  \includegraphics[height=0.20\linewidth,trim={1mm 8mm 15mm 22mm},clip]{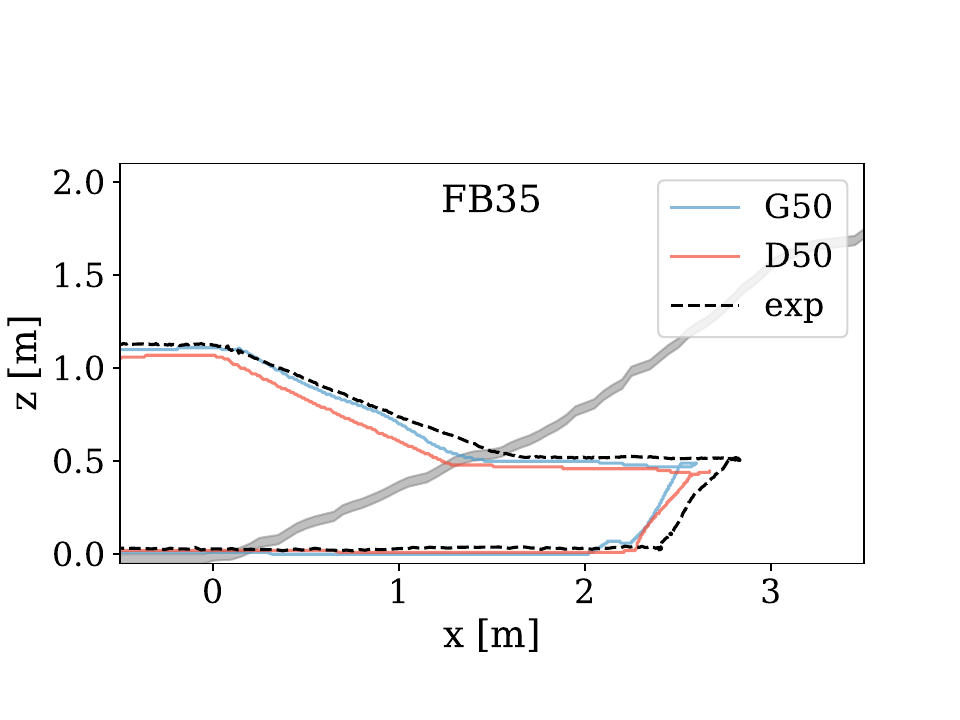}  \hspace{-2.1mm}
  \includegraphics[height=0.20\linewidth,trim={20mm 8mm 15mm 22mm},clip]{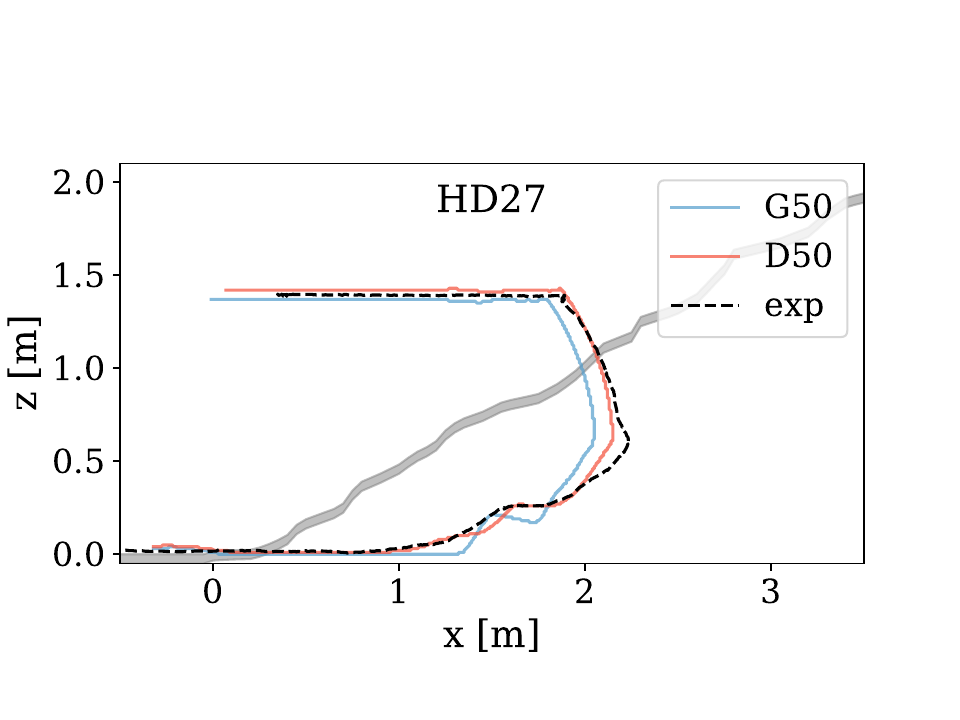}
  \hspace{-2.1mm}
  \includegraphics[height=0.20\linewidth,trim={20mm 8mm 15mm 22mm},clip]{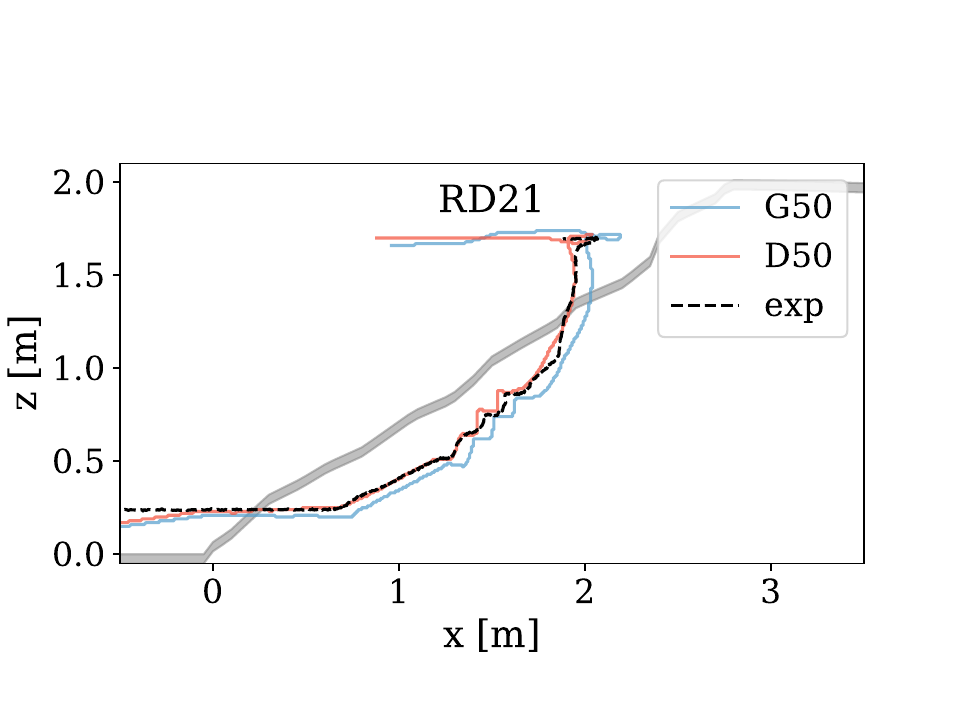} \\
  \vspace{-6.3mm}
  \includegraphics[height=0.20\linewidth,trim={1mm 8mm 15mm 22mm},clip]{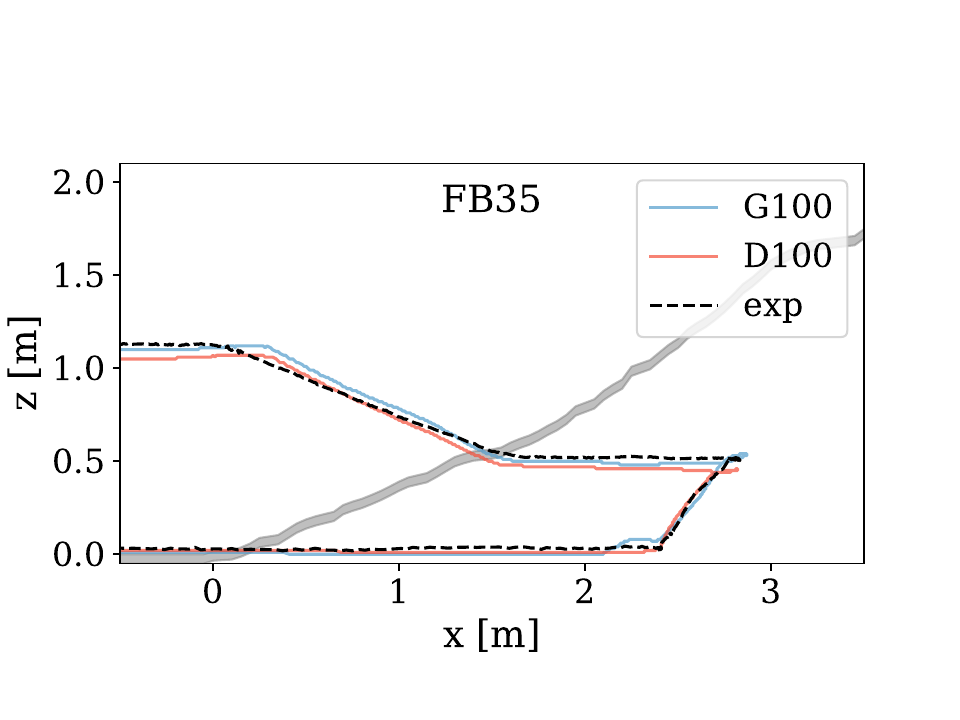}
  \hspace{-2.1mm}
  \includegraphics[height=0.20\linewidth,trim={20mm 8mm 15mm 22mm},clip]{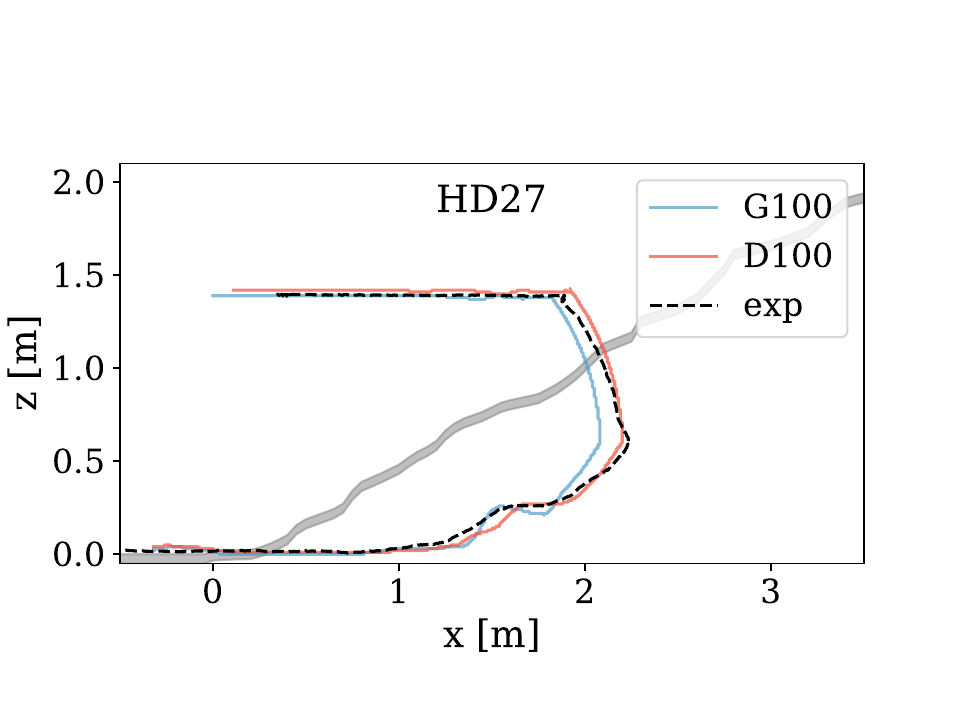}
  \hspace{-2.1mm}
  \includegraphics[height=0.20\linewidth,trim={20mm 8mm 15mm 22mm},clip]{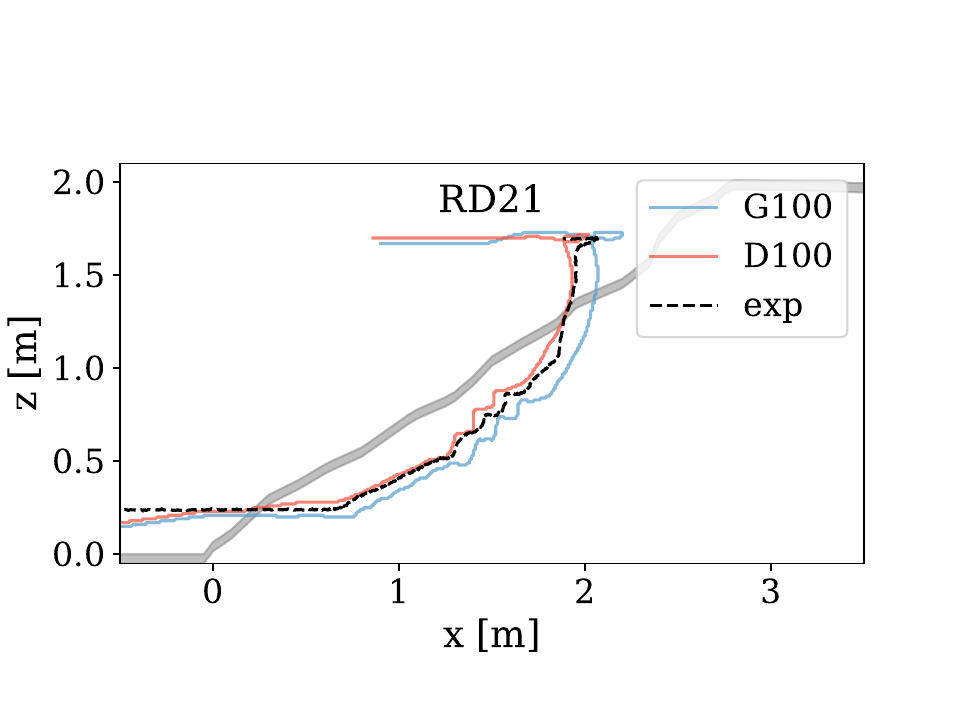} \\
  \vspace{-6.3mm}
  \includegraphics[height=0.20\linewidth,trim={1mm 8mm 15mm 22mm},clip]{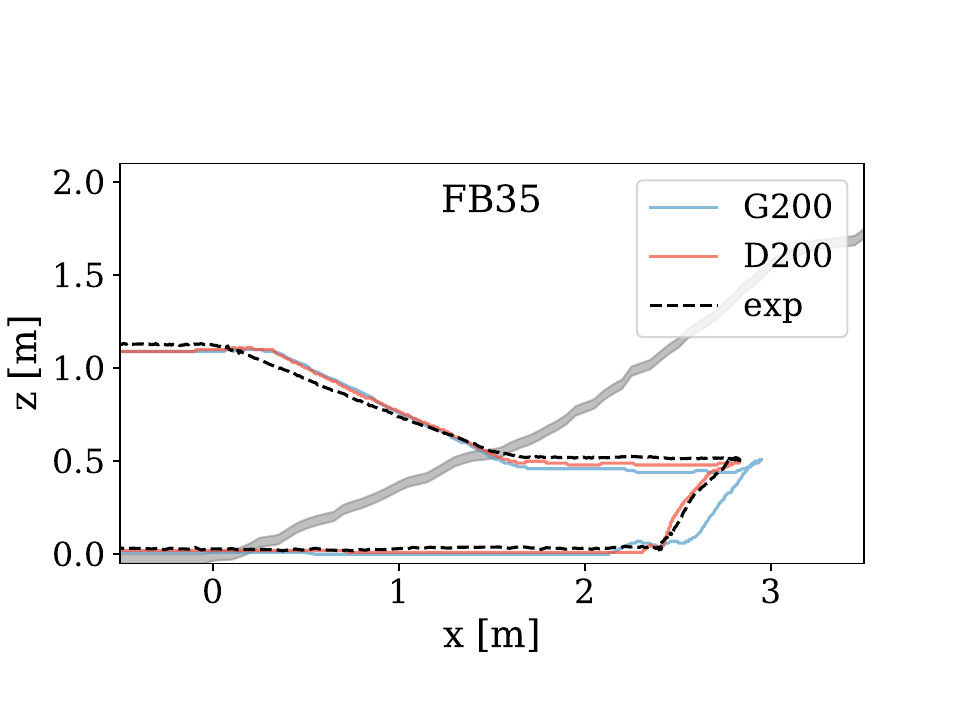}
  \hspace{-2.1mm}
  \includegraphics[height=0.20\linewidth,trim={20mm 8mm 15mm 22mm},clip]{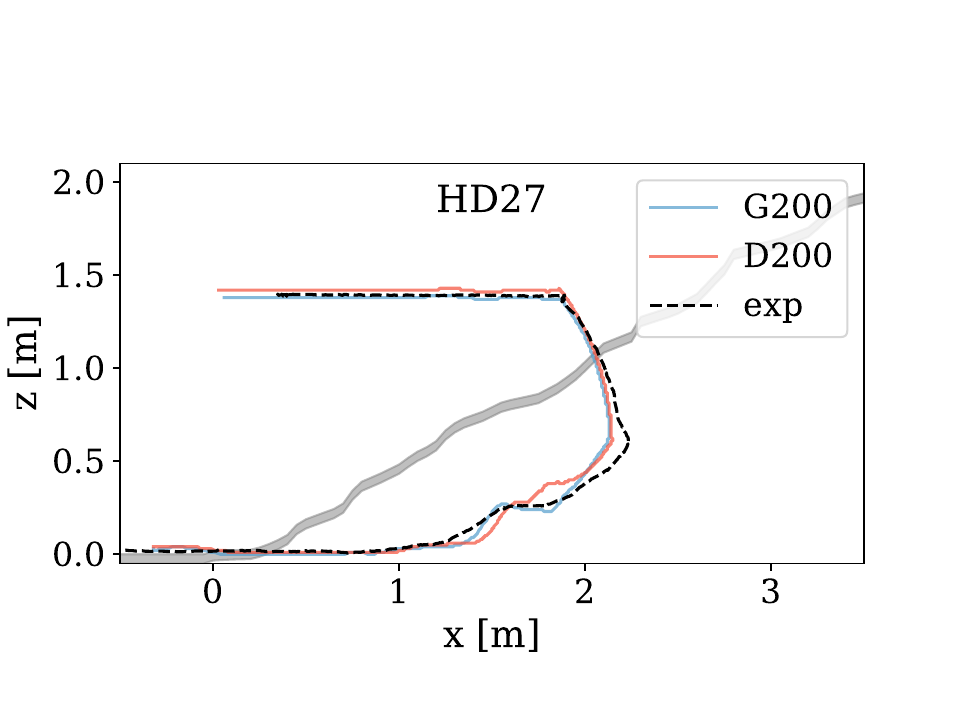}
  \hspace{-2.1mm}
  \includegraphics[height=0.20\linewidth,trim={20mm 8mm 15mm 22mm},clip]{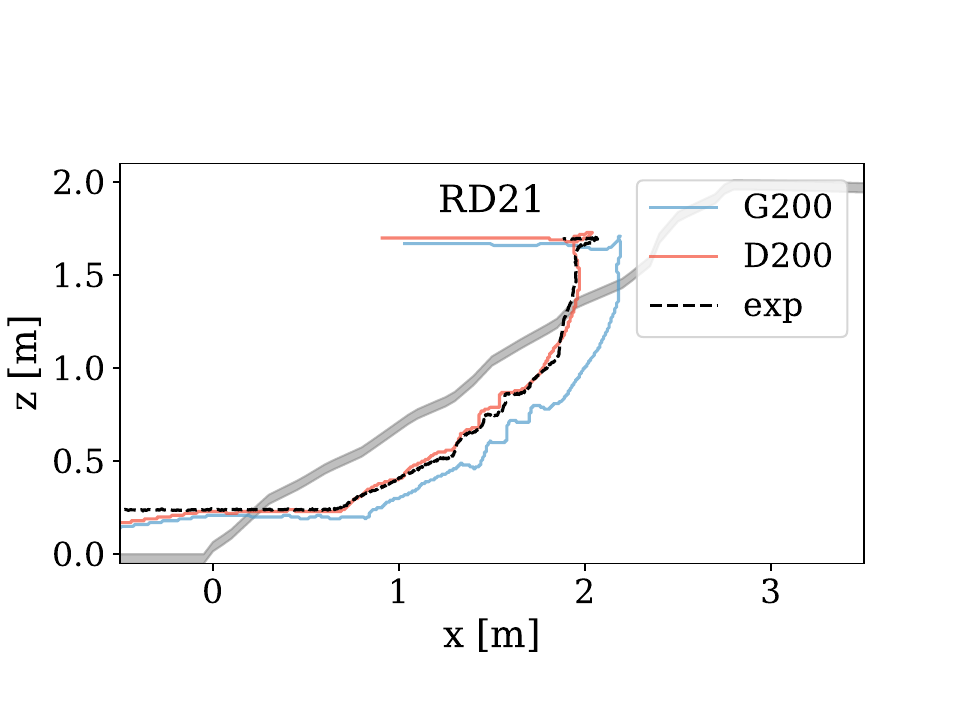} \\
  \vspace{-6.3mm}
  \includegraphics[height=0.20\linewidth,trim={1mm 8mm 15mm 22mm},clip]{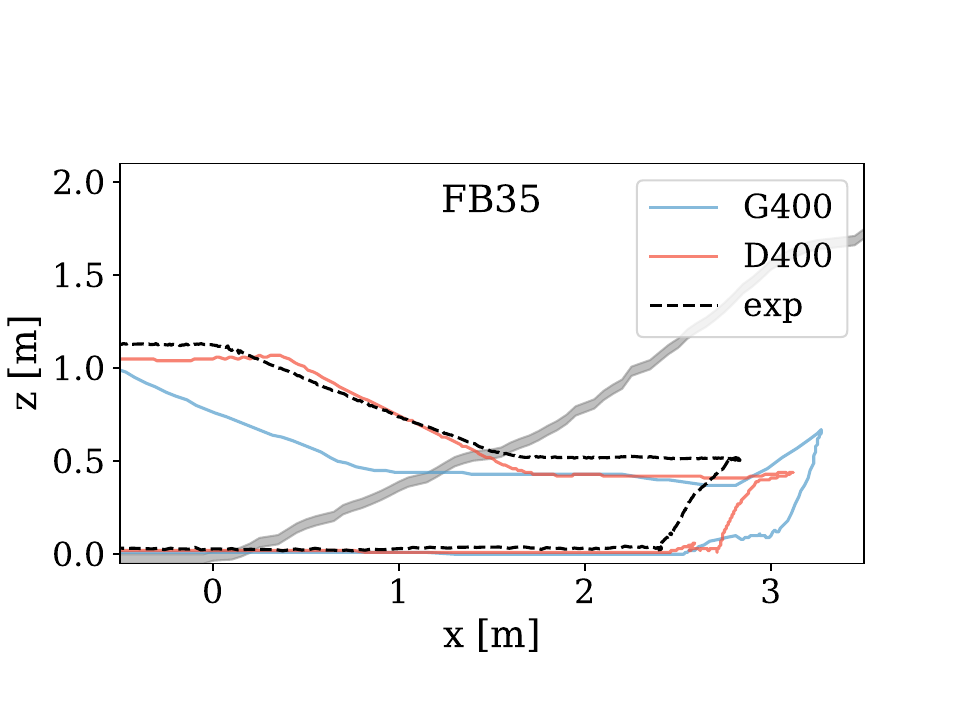}
  \hspace{-2.1mm}
  \includegraphics[height=0.20\linewidth,trim={20mm 8mm 15mm 22mm},clip]{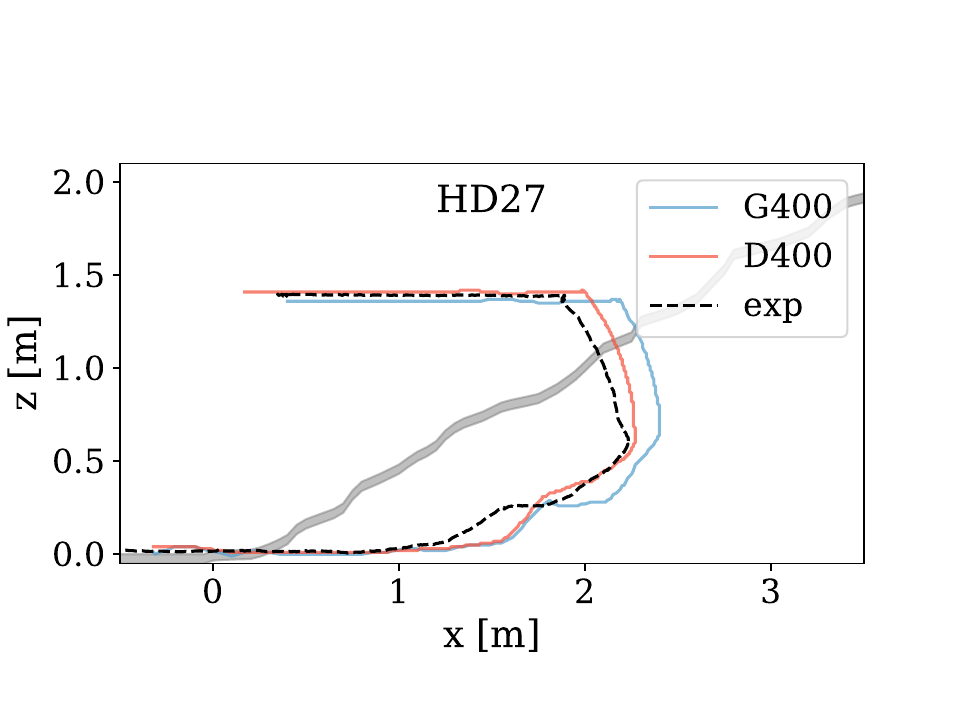}
  \hspace{-2.1mm}
  \includegraphics[height=0.20\linewidth,trim={20mm 8mm 15mm 22mm},clip]{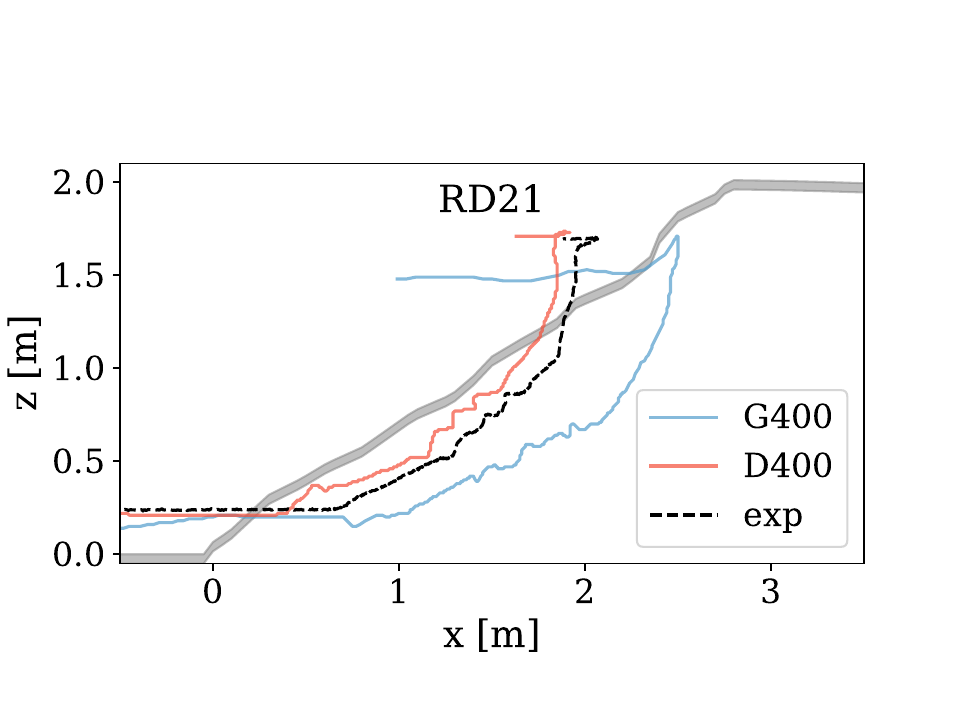} \\
  \caption{Bucket tip trajectories and initial pile shape (gray) from the field test (dashed black),
  and the simulators of type G (blue) and D (red).}
  \label{fig:trajectories}
\end{figure}

Most of the simulated bucket tip trajectories in Fig.~\ref{fig:trajectories} match the experimental ones fairly well. 
The DTW error ($\mathcal{E}_{\bm{x}}$ in Table \ref{table:sim-to-real_error_granular} 
and \ref{table:sim-to-real_error_terrain}) is, on average, 0.04 for the D-simulators and 0.09 for the G simulators.
The general trend is that the bucket penetrates too deeply with the coarsest resolution (D400 and G400), presumably because of large contact overlaps due to the large 
																							
timestep. The exception is the RD21-D400 case, where the bucket is severely obstructed from penetrating the pile surface because of the oversized particles interlocking.  The reason why this is not a problem in FB35-D400
and HD27-D400 is that the bucket penetrates along the ground plane on which the
particles rest. If the bucket had been raised half a particle diameter, the bucket would have been obstructed similarly to the RD21-D400.  
										 
There are no clear signs of excessive penetration resistance for the finer particle piles, D50\--D200.  
By design, the type-G simulators do not have this sensitivity to spatial discretization. 
The bucket's
cutting edge induces a failure plane (Fig.~\ref{fig:multiscale}) wherever it occurs 
and the penetration resistance does not depend on the location and resolution of the voxel grid. 

\subsection{Scalar time series}
\label{sec:scalar_time series}

\begin{figure}[!htb]
  \centering
  \includegraphics[width=0.8\linewidth,trim={1mm 1mm 1mm 1mm},clip]{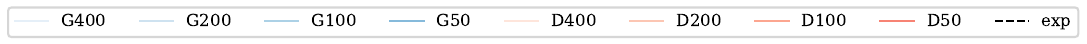}\\
  \includegraphics[height=1.0\linewidth,trim={3mm 3mm 3mm 10mm},clip]{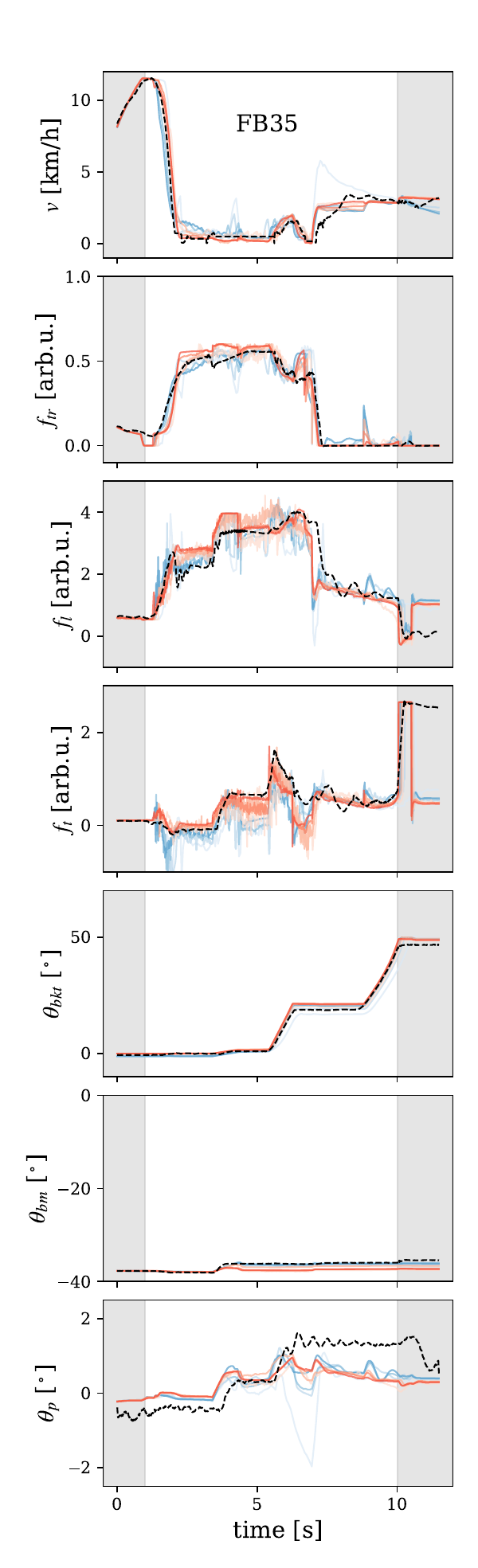} 
  \hspace{-2.5mm}
  \includegraphics[height=1.0\linewidth,trim={21mm 3mm 3mm 10mm},clip]{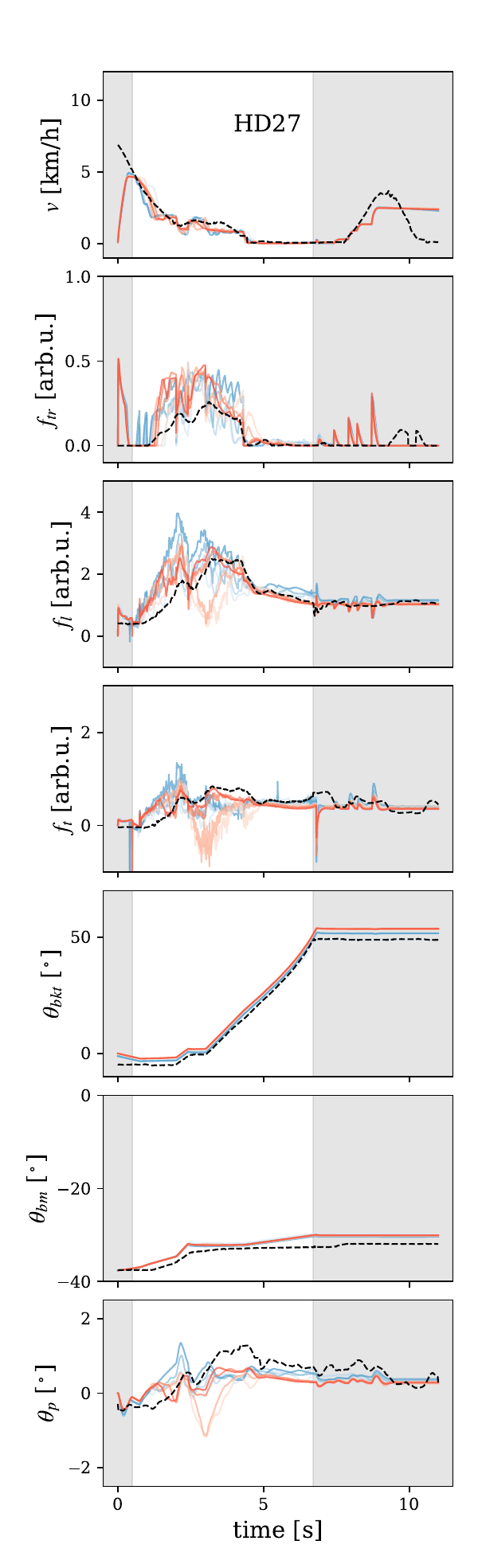}
  \hspace{-2.5mm}
  \includegraphics[height=1.0\linewidth,trim={21mm 3mm 3mm 10mm},clip]{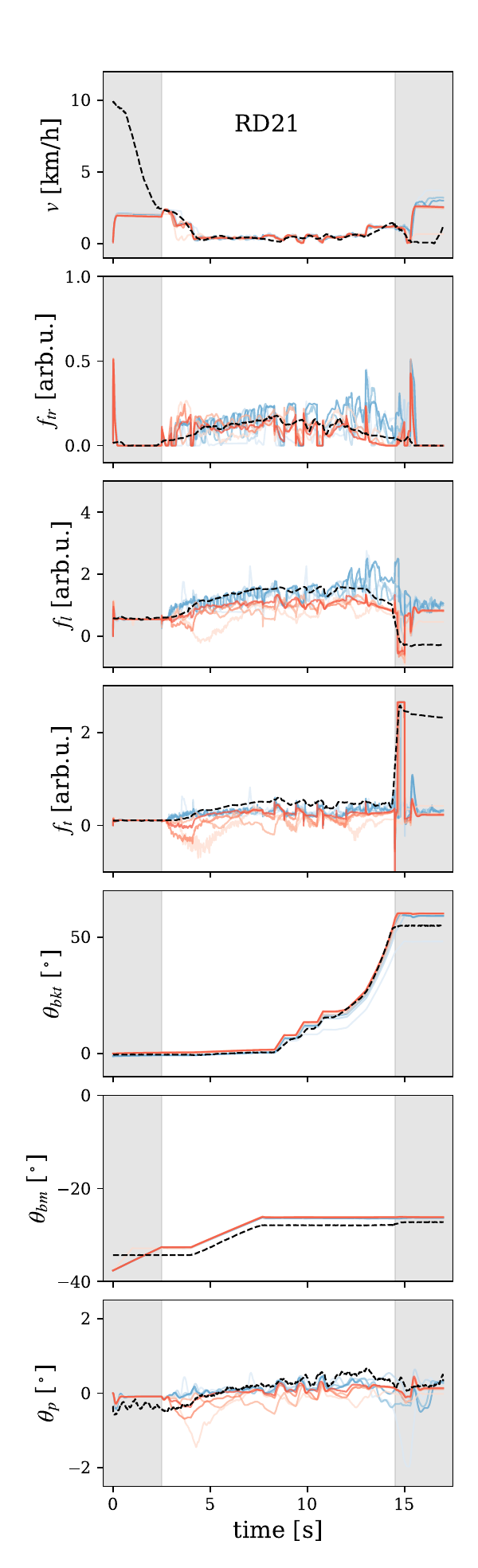}
  \caption{Speed, force, rotation measurements for time series from the FB35, HD27,
  and RD21 experiments (dashed black) and type G (blue) and D (red) simulators.
  The comparison is made in the non-shaded regions, excluding the phases of initialization and reversal after breakout. Individual plots are found in Appendix~\ref{sec:supplemental_fig}.}
     \label{fig:force_velocity}
\end{figure}

Studying the scalar time series measurements, shown in Fig.~\ref{fig:force_velocity},
we observe that the drive velocity and the rotation of the bucket and boom show
a fair agreement between simulation and experiment.
The G400 simulator stands out with the largest deviations.  The traction force MAE ($\mathcal{E}_\mathrm{tr}$ in Table \ref{table:sim-to-real_error_granular} and \ref{table:sim-to-real_error_terrain}) ranges 
between 8\% and 19\% with the largest errors when the drive, boom, and bucket 
are actuated simultaneously (HD27 and RD21). The trend of the simulated 
boom lift and bucket tilt forces match the experimentally measured one, but
there are occasionally significant deviations. The lift and tilt forces 
deviate the most during breakout in the
test FB35 around $7$ s and RD21 around $14$ s but not in the HD27 test
where breakout occurs around the $8$ s. After the breakout, 
the simulated and real lift forces are in good agreement, indicating a
good agreement in bucket filling until the time when the bucket reaches its mechanical end-point and
forces are redistributed. This happens around the $10$ s for FB35 and $14.5$ for RD21.

On average, the error in lift and tilt forces ($\mathcal{E}_\mathrm{l}$ and $\mathcal{E}_\mathrm{t}$ in Table \ref{table:sim-to-real_error_granular} and \ref{table:sim-to-real_error_terrain}) are on average 11\% and slightly smaller for the D simulators
than for the G simulators, apart from the case of D400 and D200. 
The chassis angle relative error is large (largest for the coarsest simulators) but small in absolute numbers.
The possible causes would be the model error of the tire pressures or not an entirely flat ground which we assume flat in simulation.

\subsection{Loaded mass and work}
\label{sec:mass_work}
\begin{figure}[!htb]
  \centering
  \includegraphics[height=0.14\linewidth,trim={3mm 15mm 3mm 3mm},clip]{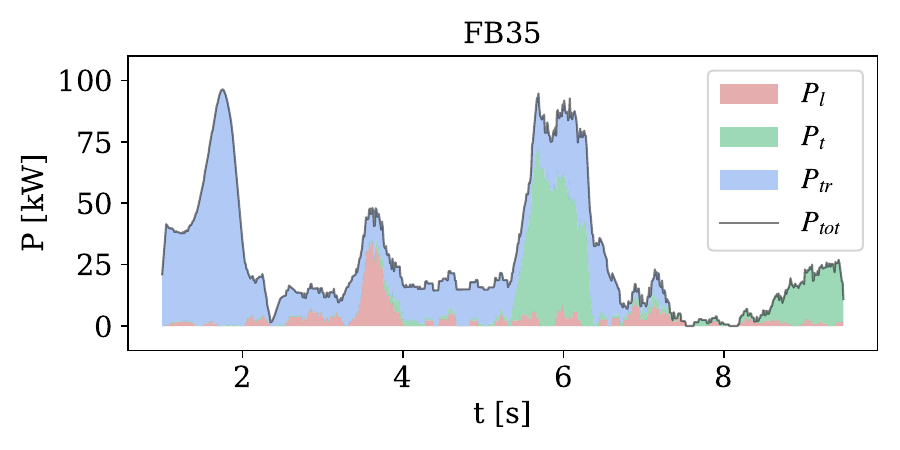}
  \hspace{-2mm}
  \includegraphics[height=0.14\linewidth,trim={21mm 15mm 3mm 3mm},clip]{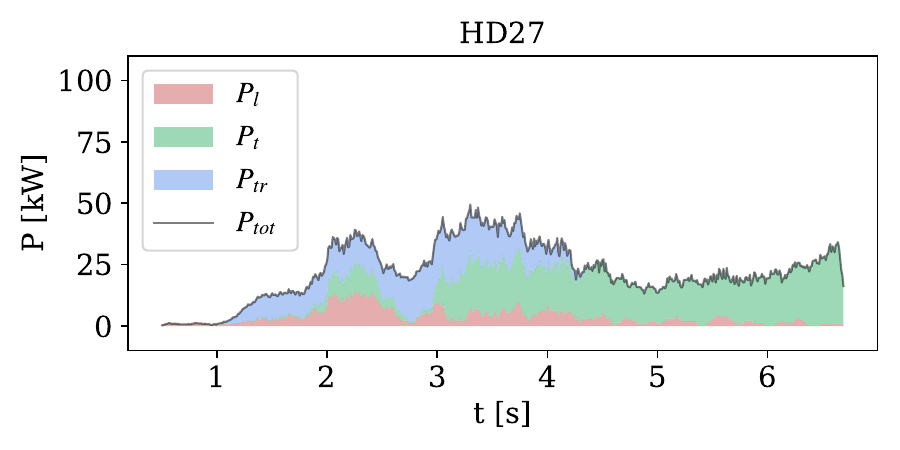}
  \hspace{-2mm}
  \includegraphics[height=0.14\linewidth,trim={21mm 15mm 3mm 3mm},clip]{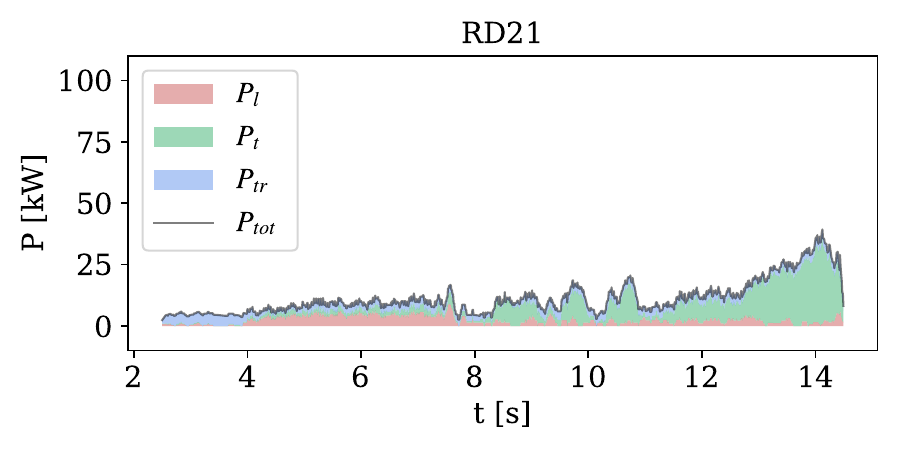}
  \put(1,33){\tiny{exp}}\\
  \includegraphics[height=0.128\linewidth,trim={3mm 15mm 3mm 8mm},clip]{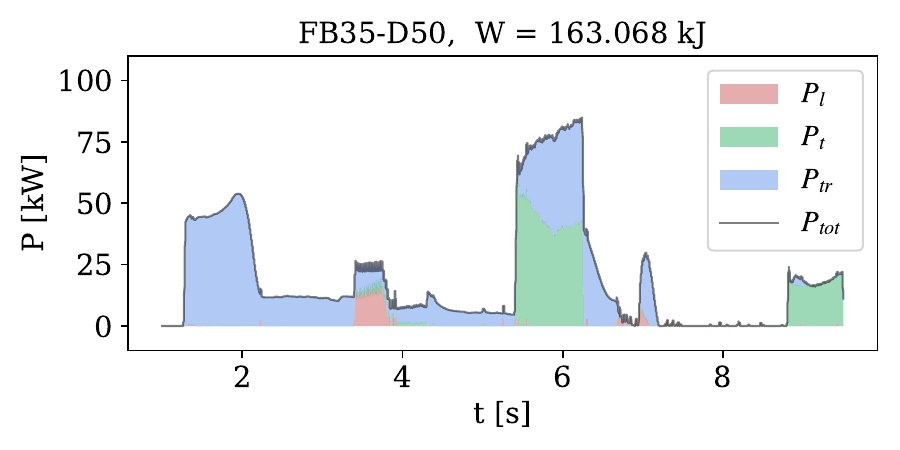}
  \hspace{-2mm}
  \includegraphics[height=0.128\linewidth,trim={21mm 15mm 4mm 8mm},clip]{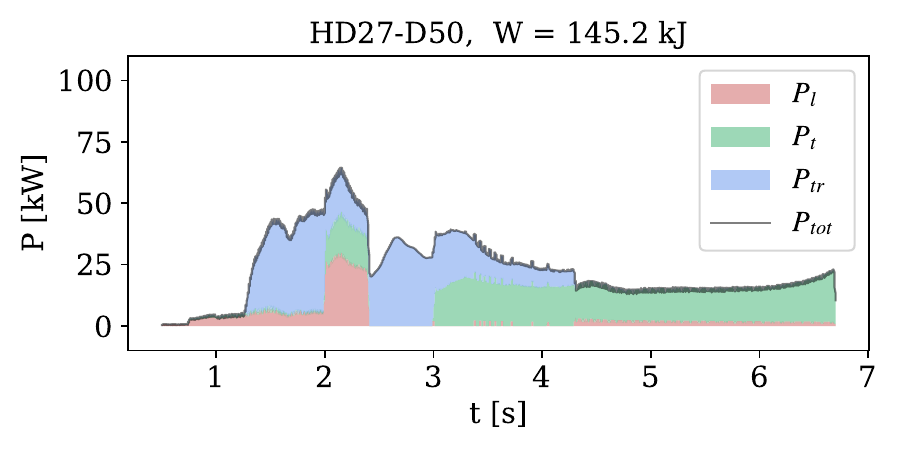}
  \hspace{-2mm}
  \includegraphics[height=0.128\linewidth,trim={21mm 15mm 3mm 8mm},clip]{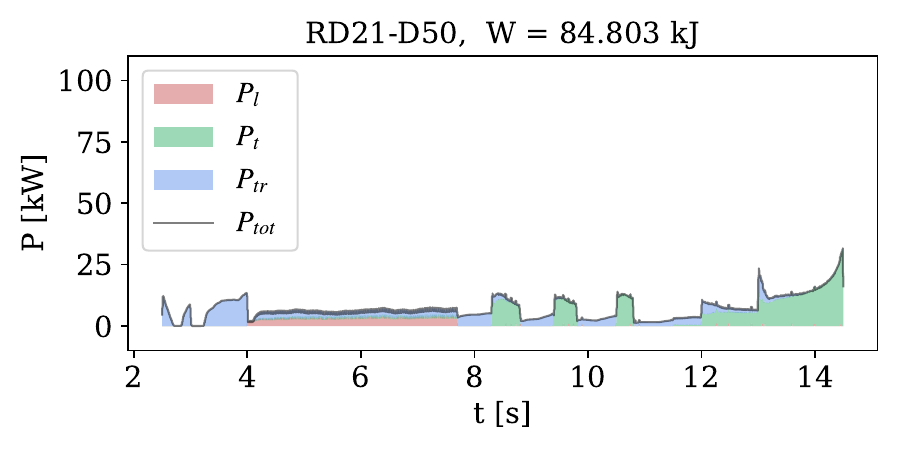}
  \put(1,33){\tiny{D50}}\\
  \includegraphics[height=0.156\linewidth,trim={3mm 3mm 3mm 8mm},clip]{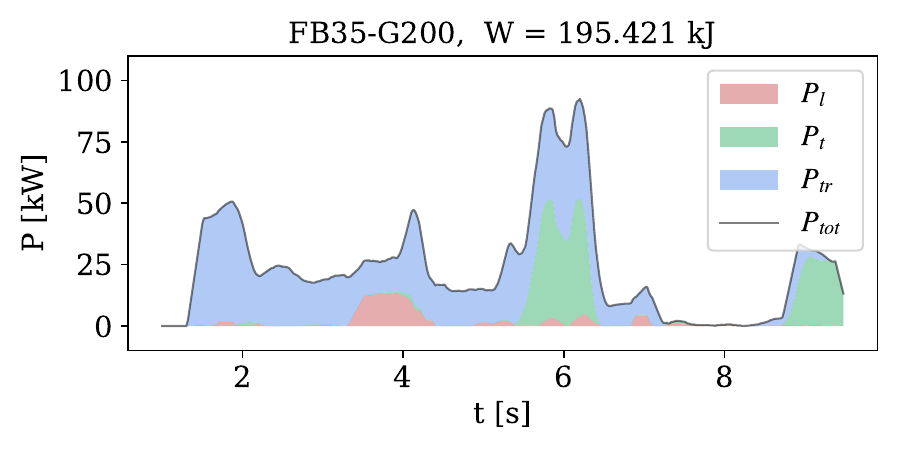}
  \hspace{-2mm}
  \includegraphics[height=0.156\linewidth,trim={21mm 3mm 3mm 8mm},clip]{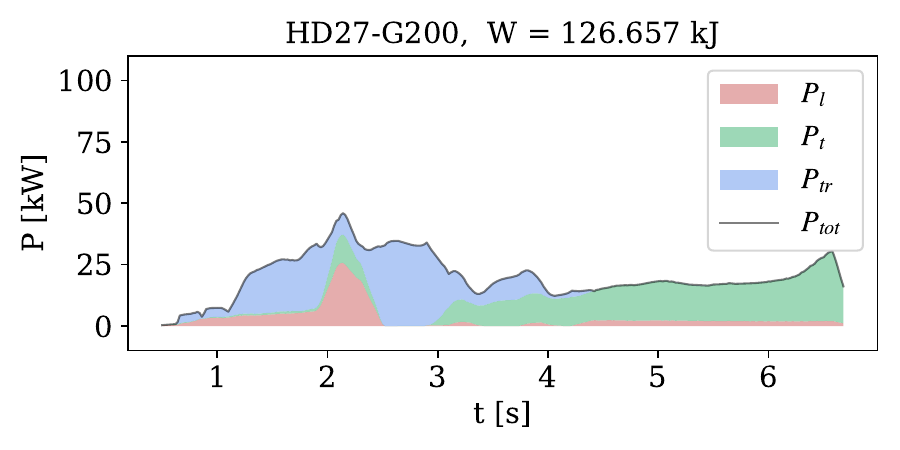}
  \hspace{-2mm}
  \includegraphics[height=0.156\linewidth,trim={21mm 3mm 3mm 8mm},clip]{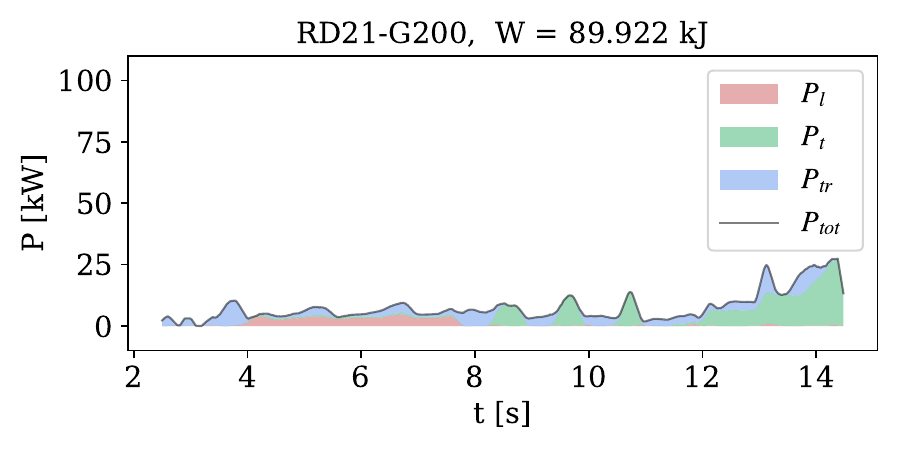}
  \put(1,42){\tiny{G200}}
  \caption{Time series of the power consumption in the three tests with sample results from the
  D50 and G200 simulators. The contributions to the total work of the drive, lift, and tilt
  actuators are shown.}
  \label{fig:work}
\end{figure}

The relative errors in loaded mass and work are listed in Table \ref{table:sim-to-real_error_granular} 
and \ref{table:sim-to-real_error_terrain}. The loaded mass is 
underestimated in the type D simulators, with a 15\% mean error, and mostly
overestimated in the G simulators, with a 12\% mean error. 
For the accumulated work, the respective mean errors are 17\% and 21\%. Again, 
the D simulators mostly underestimate the work, while G simulators mostly overestimate it. 
Sample time series of the power consumption is shown in Fig.~\ref{fig:work} with the respective contributions of
the drive, lift, and tilt actuation. Most power is consumed by driving and 
secondly by tilting. Both types of simulators show similar trends in power consumption as the field test. 

\subsection{Sim-to-real error and simulation speed}
\label{sec:sim-to-real-error}
\begin{figure}[!htb]
  \centering
  \includegraphics[width=0.5\linewidth,trim={-4mm 94mm 0mm 0mm},clip]{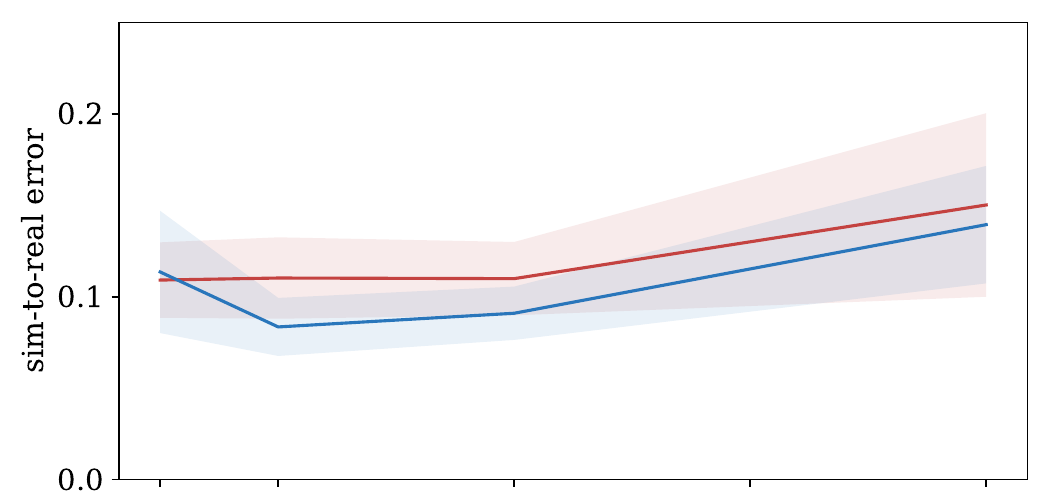}\\
  \includegraphics[width=0.5\linewidth,trim={-4mm 94mm 0mm 3mm},clip]{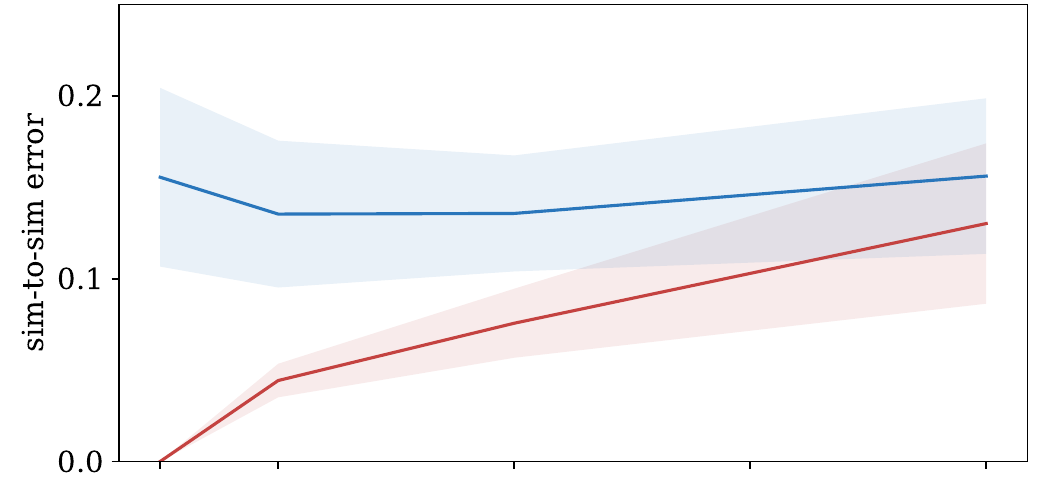}\\
  \includegraphics[width=0.5\linewidth,trim={0mm 3mm 0mm 85mm},clip]{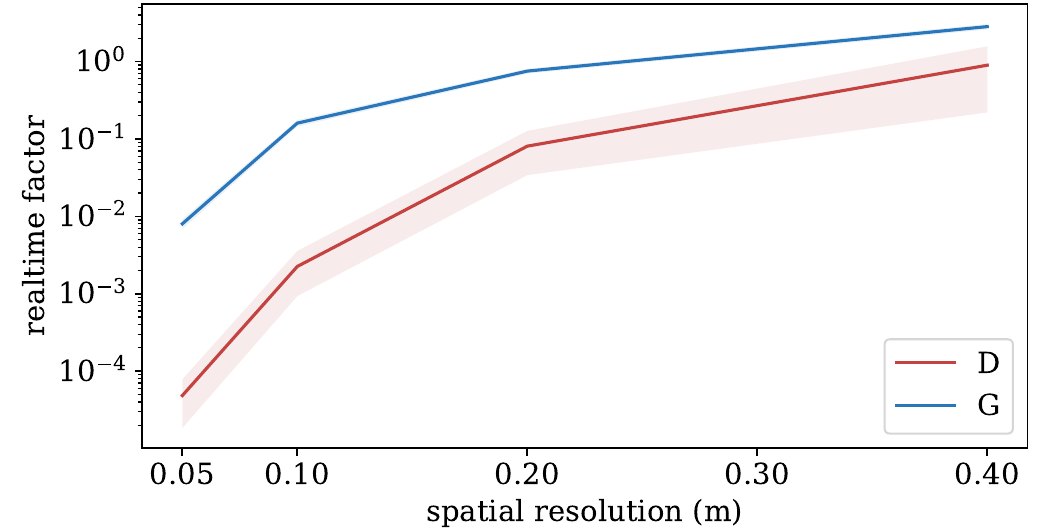}
  \caption{Sim-to-real, sim-to-sim error, and the real-time factor
  for the different levels of simulator fidelity. The solid line is the average over the three tests, and the shaded region shows the standard deviation. The spatial resolution refers to particle size and grid cell size.}
  \label{fig:error_realtimfactor}
\end{figure}

The mean error for each test and simulator fidelity was computed. 
These are listed in the right-most column in Table \ref{table:sim-to-real_error_granular} and \ref{table:sim-to-real_error_terrain}
and plotted in Fig.~\ref{fig:error_realtimfactor} (a).
We refer to this as the \emph{sim-to-real error} as it is intended to capture the reality gap of the simulators.
Overall, the sim-to-real error is
about 10\% with a standard deviation of 3\%,
On average, the sim-to-real error increases with the resolution (grid and particle size). The error is somewhat smaller for the type-G simulators than for the 
type-D simulators, except for the case of G50.

If we compare each simulator not with the field test but with the simulator of 
highest fidelity, D50, we obtain the mean 
\emph{sim-to-sim error} in Fig.\ref{fig:error_realtimfactor} (b). 
The sim-to-sim error of the D simulators increases with particle size, as can be 
anticipated since this is a self-consistency error. The G50-G400 simulators, on the other hand, are offset to D50 by a 15\% error on average.

The simulators are very different in computational intensity and speed,
as indicated by the different timestep, number of particles, and solver
iterations listed in Table \ref{table:simulators}.  
The real-time factor (computational time over simulated time) was
measured using a workstation with a single Intel i7-8700K 3.70 GHz processor.
The result is shown in Fig.\ref{fig:error_realtimfactor} (c).
The type-G simulators are roughly 100 times faster than the type-D simulators of the same resolution and run in real-time for G200 and five times faster for G400.
The G200 simulator may be considered a sweet spot in the trade-off in sim-to-real error versus speed.

\section{Domain sensitivity and predictivity with force-based control}
\label{sec:domain_sensitivity}
In this section, we investigate the domain sensitivity of a controller for automatic bucket-filling.
Imagine a controller with some free parameters, $\bm{a} \in \mathcal{A}$, that may be tuned 
for near-optimal performance by exploring the control parameter space $\mathcal{A}$
using simulations. It is then of interest how sensitive the choice of control parameters is 
under transfer to the target domain. In other words, is a control parameter that is found to be
near-optimal in simulation also near-optimal in reality? We did not have the possibility of 
running additional field experiments. Instead, we examined the domain sensitivity of a 
controller optimized with the fast G200 simulator under transfer to the D50 simulator,
which is much more finely resolved and slower than four orders in magnitude. From Fig.~\ref{fig:error_realtimfactor},
we know that the gap between G200 and D50 is similar in size to the simulation-to-reality gap,
albeit the nature of deviations is different. 

\subsection{Test setup} 
We used the same force feedback controller for automatic bucket filling as studied in 
\cite{aoshima:2023:pmh} and run it on the FB35 test pile. The wheel loader starts 5 m from the pile, heading straight
with the target speed of 8 km/h and with the bucket lowered horizontally to the ground.
Once the bucket reaches the pile, the force feedback control law is engaged on the lift 
and bucket cylinders to fill the bucket until the bucket tip breaks out from the pile.
After breakout, the machine is held still for 0.5 s, for the soil to settle and then
starts reversing with a target speed of 8 km/h while lifting and tilting to reach the final boom 
and bucket angles of $-20^\circ$ and $50^\circ$, respectively. The loading cycle ends when the machine 
reaches the starting point. 

The force feedback controller is an adaptation of the 
admittance controller in \cite{Dobson2017}. It determines the target speed of the linear motors for the boom and 
bucket cylinders. Recall that the linear motors are modeled as velocity constraints with force range limits on the constraint force, hence the set target speed will not be realized if the required force is not within the motor limits. The controller is defined by the following target speeds: $v_\mathrm{bm}^\text{target} =  u_\mathrm{bm}(f_\mathrm{bm},\bm{a}) v_\mathrm{bm}^\text{max}$ 
and $v_\mathrm{bk}^\text{target} =  u_\mathrm{bk}(f_\mathrm{bm},\bm{a}) v_\mathrm{bk}^\text{max}$, where 
$f_\mathrm{bm}$ is the measured force in the boom cylinder, suitably normalized.
The response functions are $u_\mathrm{bm} = \text{clip}\left( k_\mathrm{bm}\left[ f_\mathrm{bm} -\delta_\mathrm{bm} \right],0,1\right)$ and $u_\mathrm{bk} = \text{clip}\left( k_\mathrm{bk}\left[ f_\mathrm{bm} -\delta_\mathrm{bk} \right],0,1\right)$, where $\text{clip}(value, min, max)$ limits $value$ to the maximum and minimum values.
The digging resistance increases with the depth of bucket penetration into the pile.
If the dig resistance, observed through the boom cylinder force $f_\mathrm{bm}$, exceeds the threshold parameters $\delta_\mathrm{bm}$ or $\delta_\mathrm{bk}$, then the lift or tilt actuation is engaged, respectively.
Larger values of the threshold parameters will typically render bucket trajectories with deeper penetration.
The gain parameters, $k_\mathrm{bm}$ and $k_\mathrm{bk}$, regulate how rapid the respective reactions are.
These are collected in a control parameter vector $\bm{a} = [\delta_\mathrm{bm}, k_\mathrm{bm}, \delta_\mathrm{bk}, k_\mathrm{bk}]$.
It should be noted that unlike the feedforward controller in Sec.~\ref{sec:comparison}, the
time for completing the loading cycle is entirely unknown and highly dependent on the control parameter
and pile shape.
 
For simplicity, we do not cover the full four-dimensional control parameter space here. Instead,
we sweep along a search line $\bm{a}(s) = \bm{a}_0 + (\bm{a}_1-\bm{a}_0)s$, 
with $\bm{a}_0 = [0.7, 0.3, 0.2, 0.2]$, $\bm{a}_1 = [0.0, 2.2, 0.15, 4.8]$, and search parameter $s \in [0,1]$.
Typically, $s \approx 0$ produces a deep bucket penetration before breaking out while $s\approx 1$
will render a shallower trajectory following the surface. 
Simulations were run with distinct control parameters by sweeping $s$ from $0.0$ to $1.0$ in 50 equally spaced
intervals for G200 and 30 intervals for D50.  To see the dependency on the simulator level-of-fidelity,
simulations with G100 and D100 were also run. Samples are shown in Supplementary Video 4.

\subsection{Resulting domain sensitivity}
\begin{figure}[!ht]
  \centering
  \begin{subfigure}{0.25\textwidth}
    \includegraphics[trim={0 0 935 0},clip, width=1.0\linewidth]{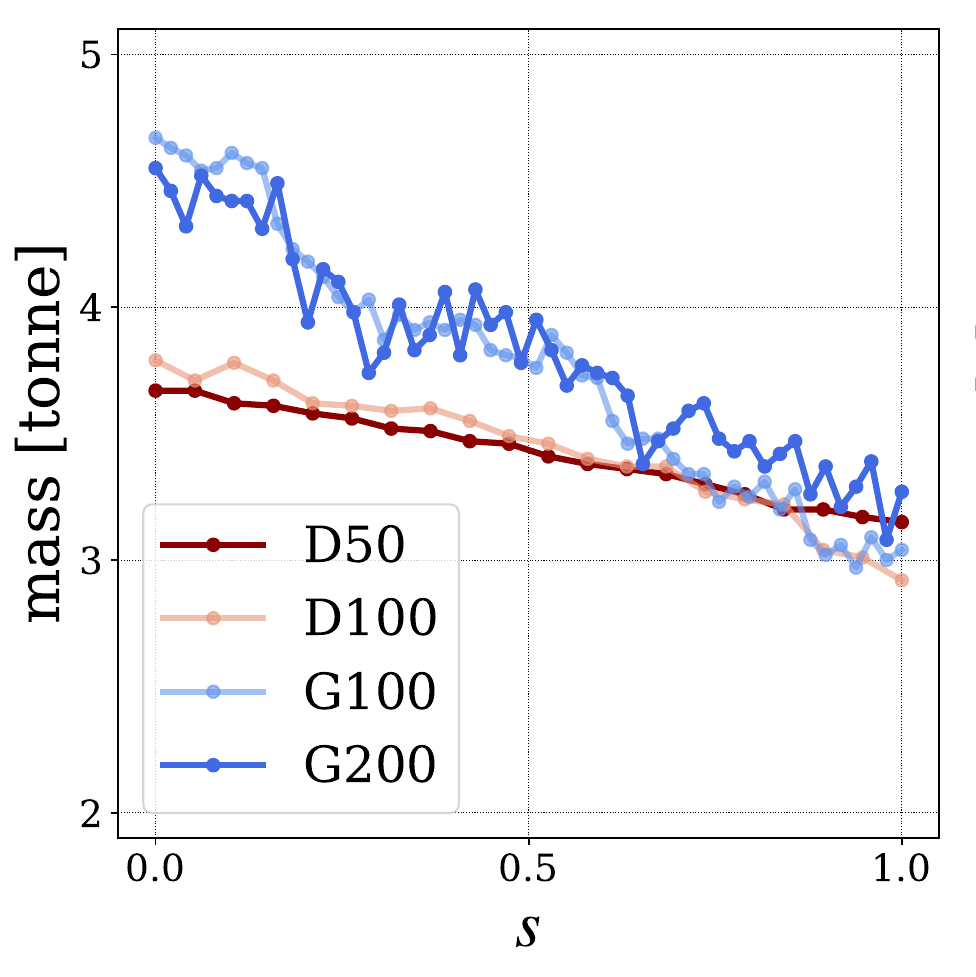}
    \caption{}
  \end{subfigure}
  \begin{subfigure}{0.25\textwidth}
    \includegraphics[trim={455 0 475 0},clip, width=1.0\linewidth]{fig/Observation_TransferTest.pdf}
    \caption{}
  \end{subfigure}
  \begin{subfigure}{0.25\textwidth}
    \includegraphics[trim={930 0 0 0},clip, width=1.0\linewidth]{fig/Observation_TransferTest.pdf}
    \caption{}
  \end{subfigure}
  \caption{The dependency of loaded mass (a), time (b), and work (c) on the force feedback control 
  parameter $s$ in the domains G200 and D50.  To see the sensitivity on resolution, G100 
  and D100 are also shown.}
  \label{fig:transfer_test_observation}
\end{figure}

The measured load mass $M$, cycle time $T$, and work $W$ for the control parameters and
simulators are shown in Fig.~\ref{fig:transfer_test_observation}. The simulators show the same general
dependency but with some differences. The mass, time, and work
are monotonically decreasing with $s$, with some fluctuations that are larger for G than for D.
The load time agrees well, capturing how time increases rapidly with deep bucket penetration.
For the work, there is a nearly constant gap of 50 kJ, G200 yielding roughly 
15\% higher values than D50. For the mass, the gap is 25\% in the region of maximal bucket filling, occurring for $s\approx 0$,
and decreases steadily with increasing $s$.  These gaps are consistent with the results in Sec.~\ref{sec:comparison},
where feedforward control was used. The dependency on resolution is hardly notable here.

\begin{figure}[ht]
  \centering
  \includegraphics[trim={0 0 0 0},clip, width=0.4\linewidth]{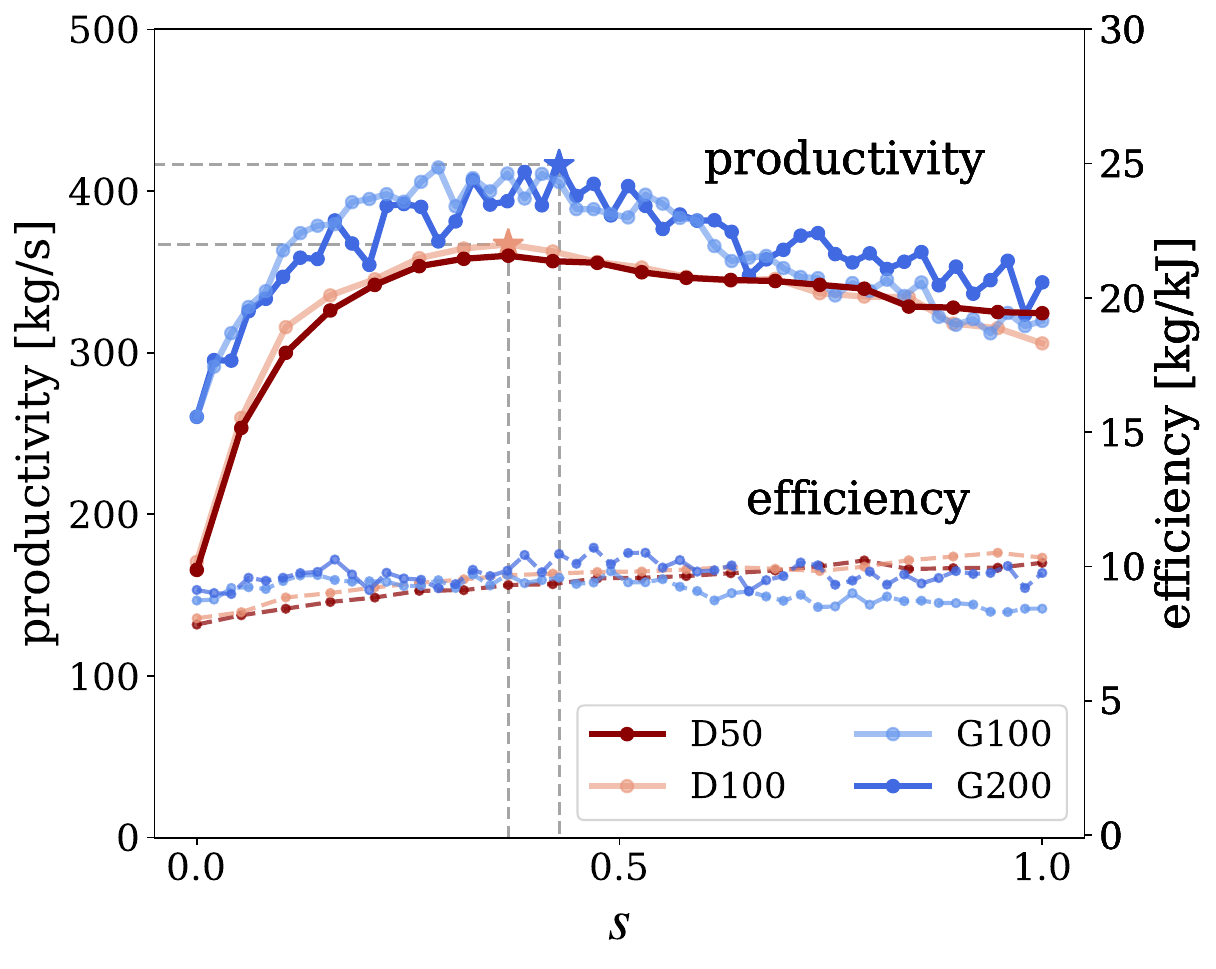}
  \caption{Domain sensitivity from force feedback control with parameter $\bm{a}(s)$ in the simulation domains G100, G200, D50, and D100.}
  \label{fig:domain_sensitivity}
\end{figure}

The productivity and efficiency of each simulated loading are computed as $M/T$ and $M/W$, respectively.
Their dependency on the control parameter and on the type of simulator is shown in Fig.~\ref{fig:domain_sensitivity}.
The absolute value in productivity differs because of the mentioned gap, but the trends are similar.
Productivity drops in a similar fashion when the load time increases sharply as $s$ approaches zero.
The D simulators predict that efficiency is monotonically increasing with $s$ while it is
more or less constant for G.

If the task was to select an optimal control parameter using the G200 simulator
and transfer it to the D50 domain, we would have experienced a domain gap
as well as a shift in the value of the optimal parameter in the different domains.
In the current example, the maximal productivity in G200 is about $416$ kg/s at $s=0.42$, while D50 simulations observed the maximal productivity of $366$ kg/s at $s=0.36$.
This translates into a domain gap of 49 kg/s (13\%) and a domain shift of 0.06.
If the control parameter $s=0.42$, which is optimal for G200, were directly transferred 
to the D50 domain, the performance drop from the found optimal value would only be 2\%.
Since this might be the special case of the selected action space, we also tested the domain sensitivity in an additional 10 spaces.
To avoid the simulation cost of D50, the additional tests were conducted between G200 and D100, assuming the gap between D50 and D100 was marginal, as shown in Fig.~\ref{fig:domain_sensitivity}.
In consequence, the average of the domain gap, the domain shift, and the performance drop resulted in 55 kg/s (16\%), 0.22, and 5\%, respectively.

\section{Discussion}
\label{sec:discussion}
A limitation of the present study is the specificity of loading homogeneous gravel. 
It is likely that the sim-to-real gap will be larger when considering more 
complex and heterogeneous soil, such as coarse fragmented rocks and cohesive dirt.
On the other hand, the results in \cite{Eriksson2023} suggest that the adaptation 
to other materials is not an insurmountable problem.

The wheel loader model in this paper is highly simplified, in particular the engine and power transmission through the driveline and hydraulics for the boom lift and bucket tilt. 
In reality, they share and compete for the same power source. A simple model extension 
that would not affect simulation timestep or speed would be to adopt the model in \cite{Servin2018}. The number of parameters to calibrate would, however, increase.

\section{Conclusion}
\label{sec:conclusion}

We found that it is possible to create a full-system wheel loading simulator with a sim-to-real
gap of 10\%. If the domain sensitivity between D and G simulators is representative
of the true reality gap, this level of sim-to-real gap is clearly sufficient to transfer the studied force feedback controller without a significant drop in optimality. 
The observed gap depends weakly on the simulated terrain's level of fidelity.
Unexpectedly, the reduced multiscale terrain model can deliver as good or better realism as a DEM model
despite several orders of magnitude differences in degrees of freedom and computational speed.
The fact that it has more free model parameters is compensated by high computational speed,
allowing for many more evaluations during calibration. The findings suggest that the 
observed simulation-to-reality gap is due more to model errors than numerical errors. 
To further reduce the gap, we advise a more refined model of the engine, the hydraulic actuation
of the boom and bucket and power distribution between it and the driveline.

\section*{Supplementary material}
\label{sec:supplementary}
Supplementary data to this article can be found online at \url{http://umit.cs.umu.se/wl-sim-to-real/}.
\begin{itemize}
    \item [] \textbf{Supplementary video 1} \\
    The field experiment and corresponding simulations for the three bucket-filling tests FB35, HD27, and RD21.
    The particle and the multiscale terrain models are D50 (top right), G200 (bottom left), and G400 (bottom right).
    \item [] \textbf{Supplementary video 2} \\
    The simulations of the eight different levels of fidelity for the bucket-filling tests FB35, HD27, and RD21.
    D50/G50 (top left), D100/G100 (top right), D200/G200 (bottom left), and D400/G400 (bottom right).
    \item [] \textbf{Supplementary video 3} \\
    The simulations of the bucket-filling tests FB35, HD27, and RD21 with overlaid images from the D50 and G200 simulators.
    Particles are color-coded by velocity.
    \item [] \textbf{Supplementary video 4} \\
    The optimized bucket-filling controller in each simulator: D50 (top left), D100 (top right), G100 (bottom left), and G200 (bottom right).
    Particles are color-coded by velocity.
\end{itemize}

\section*{Appendix - Supplementary figures}
\label{sec:supplemental_fig}
\begin{figure}
  \centering
  \includegraphics[height=1.0\linewidth,trim={3mm 3mm 3mm 10mm},clip]{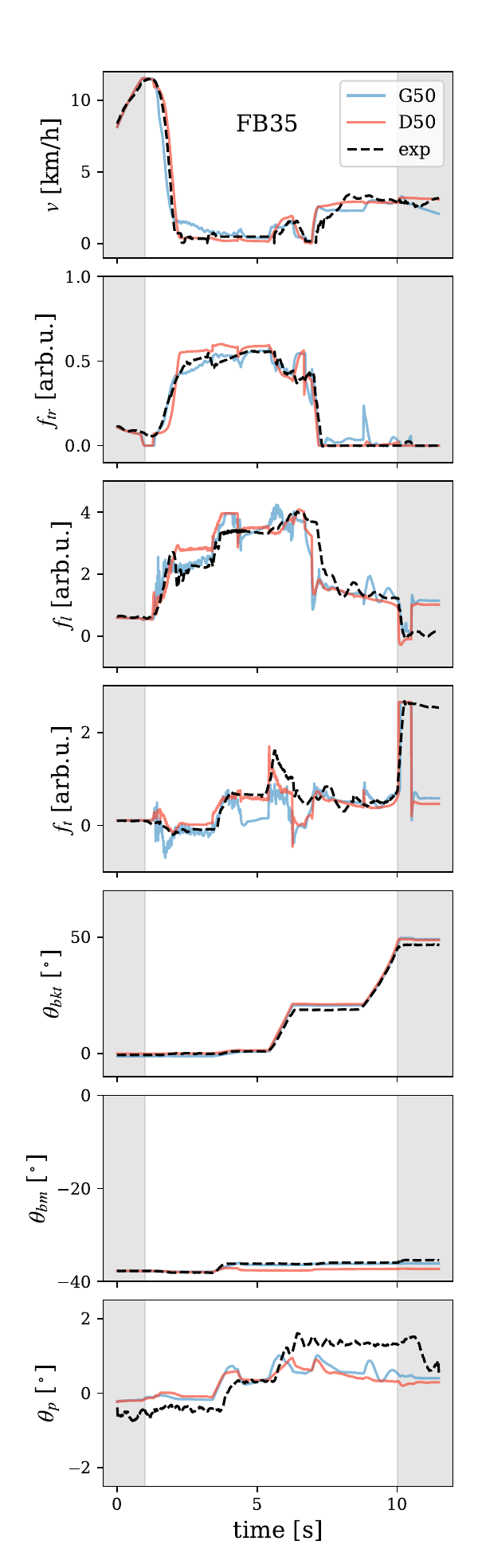} 
  \hspace{-2.5mm}
  \includegraphics[height=1.0\linewidth,trim={19mm 3mm 3mm 10mm},clip]{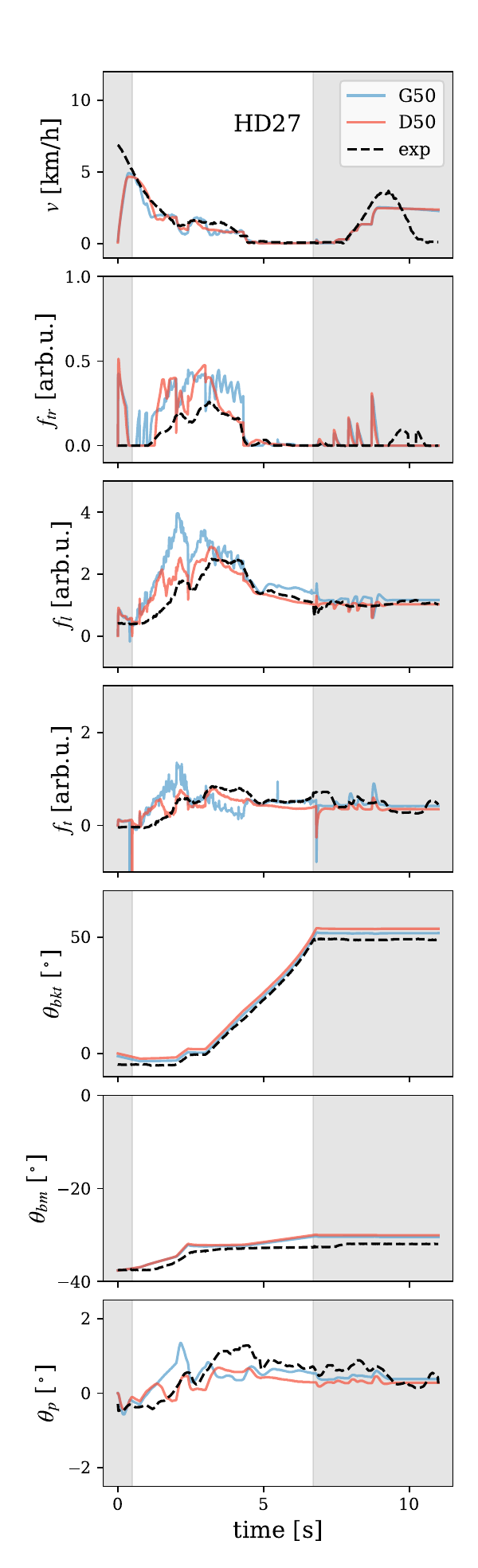}
  \hspace{-2.5mm}
  \includegraphics[height=1.0\linewidth,trim={19mm 3mm 3mm 10mm},clip]{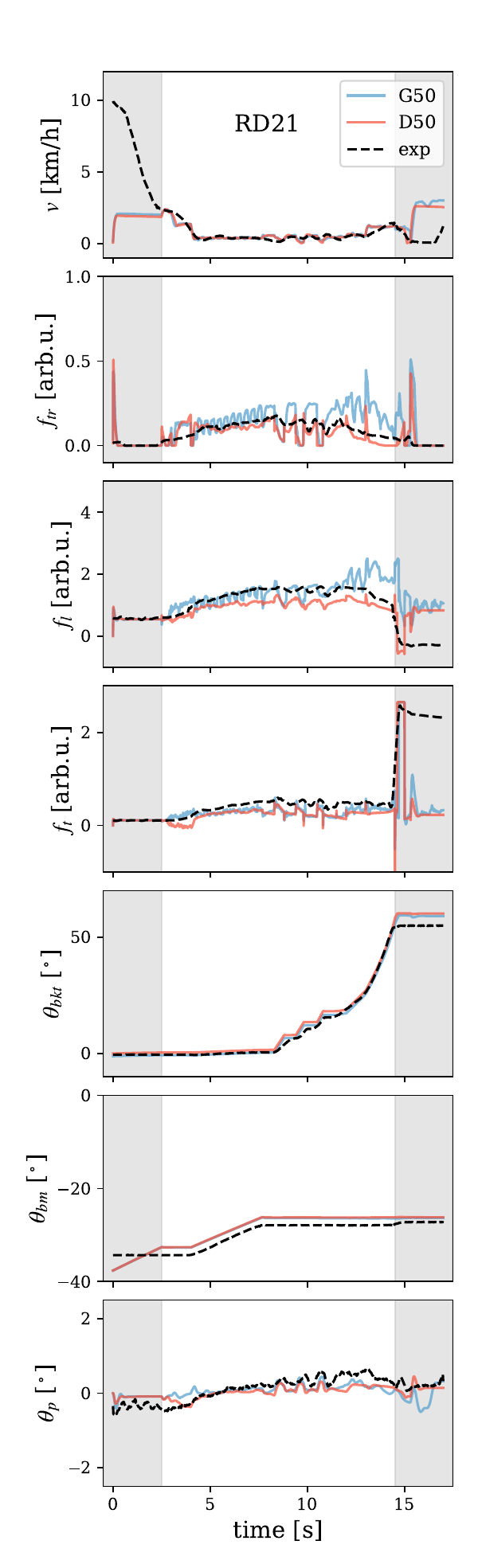}
  \caption{Speed, force, rotation measurements for time series from the FB35, HD27, RD21 experiments and type G50 and D50 simulator results.}
  \label{fig:suppl-GD50}
\end{figure}

\begin{figure}
  \centering
  \includegraphics[height=1.0\linewidth,trim={3mm 3mm 3mm 10mm},clip]{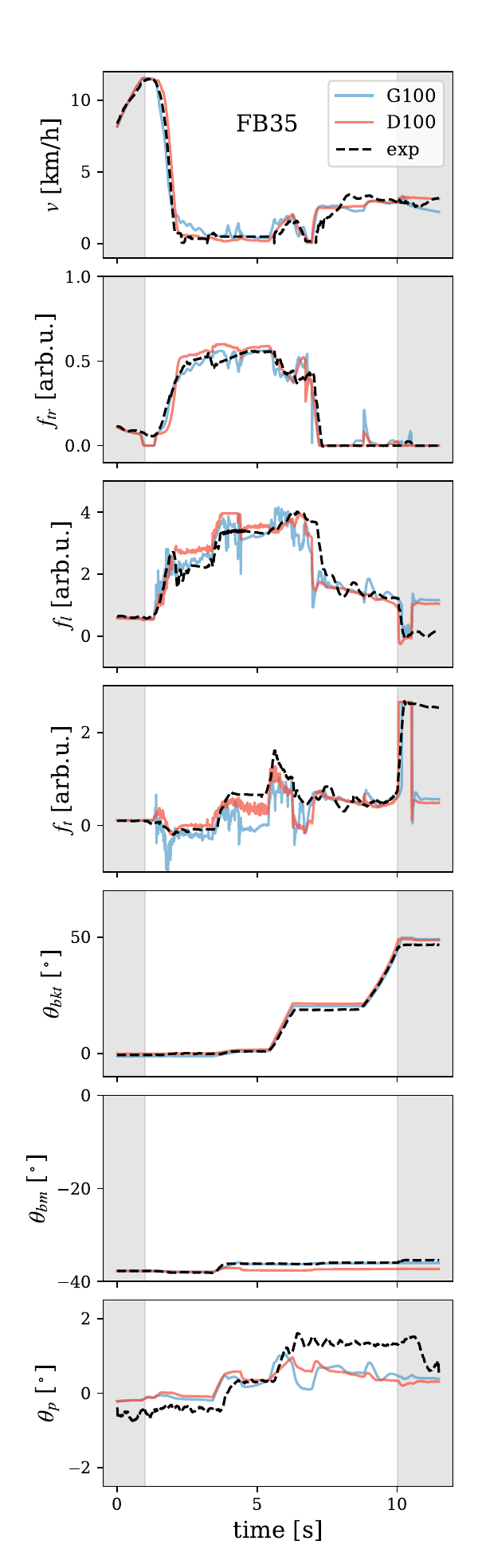} 
  \hspace{-2.5mm}
  \includegraphics[height=1.0\linewidth,trim={19mm 3mm 3mm 10mm},clip]{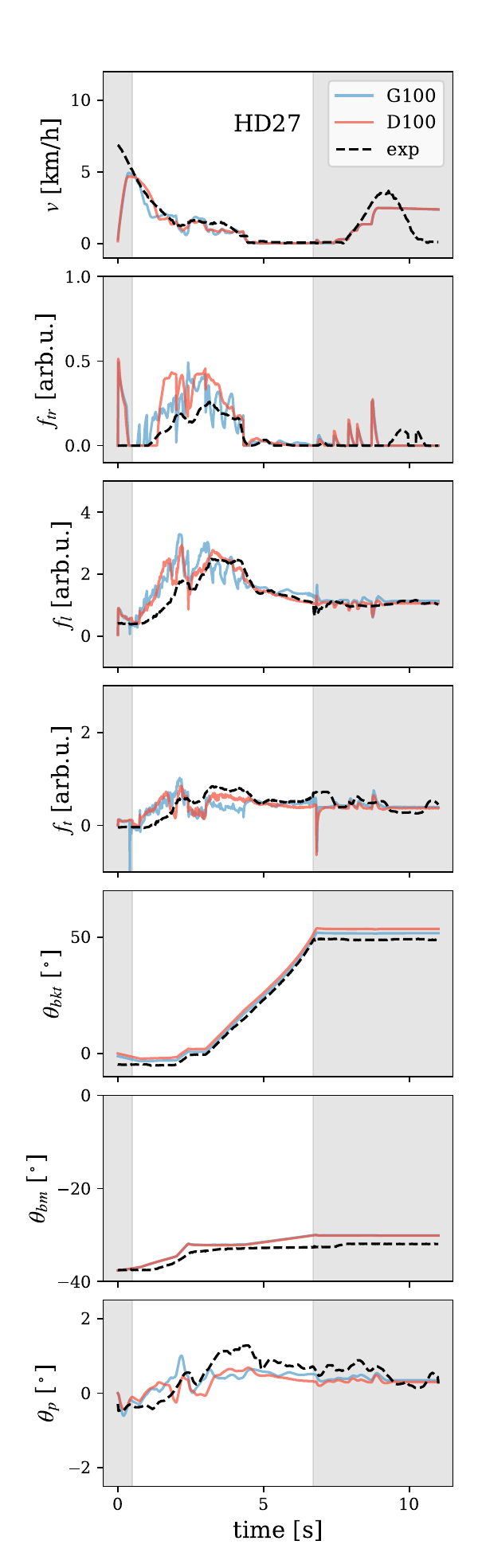}
  \hspace{-2.5mm}
  \includegraphics[height=1.0\linewidth,trim={19mm 3mm 3mm 10mm},clip]{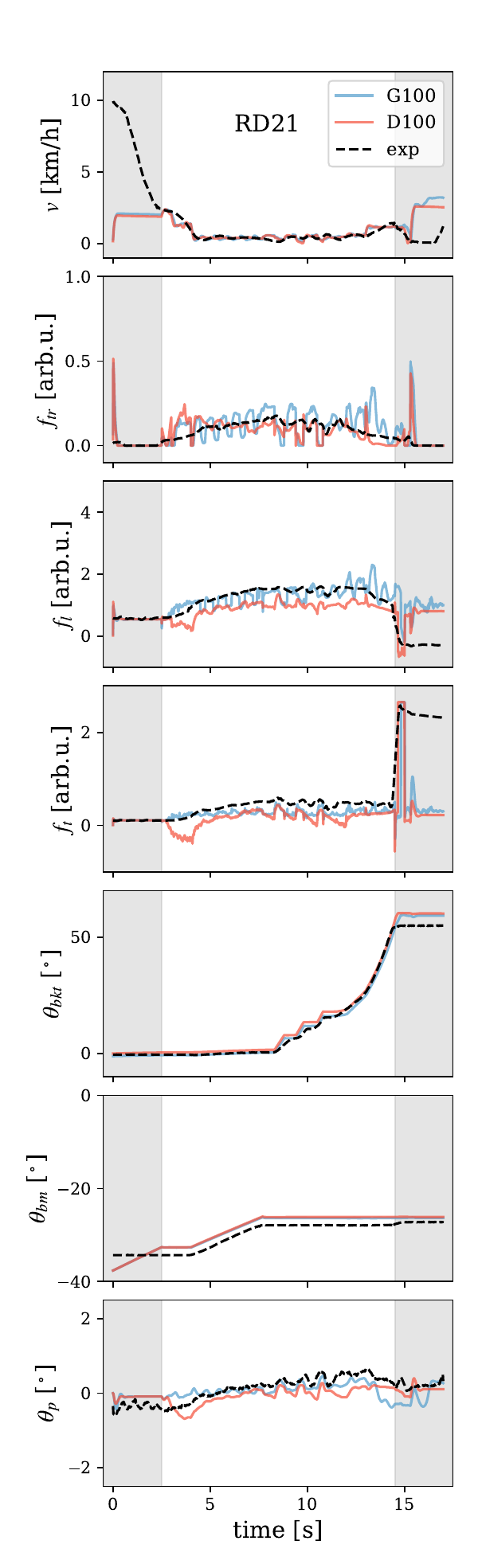}
  \caption{Speed, force, rotation measurements for time series from the FB35, HD27, RD21 experiments and type G100 and D100 simulator results.}
  \label{fig:suppl-GD100}
\end{figure}

\begin{figure}
  \centering
  \includegraphics[height=1.0\linewidth,trim={3mm 3mm 3mm 10mm},clip]{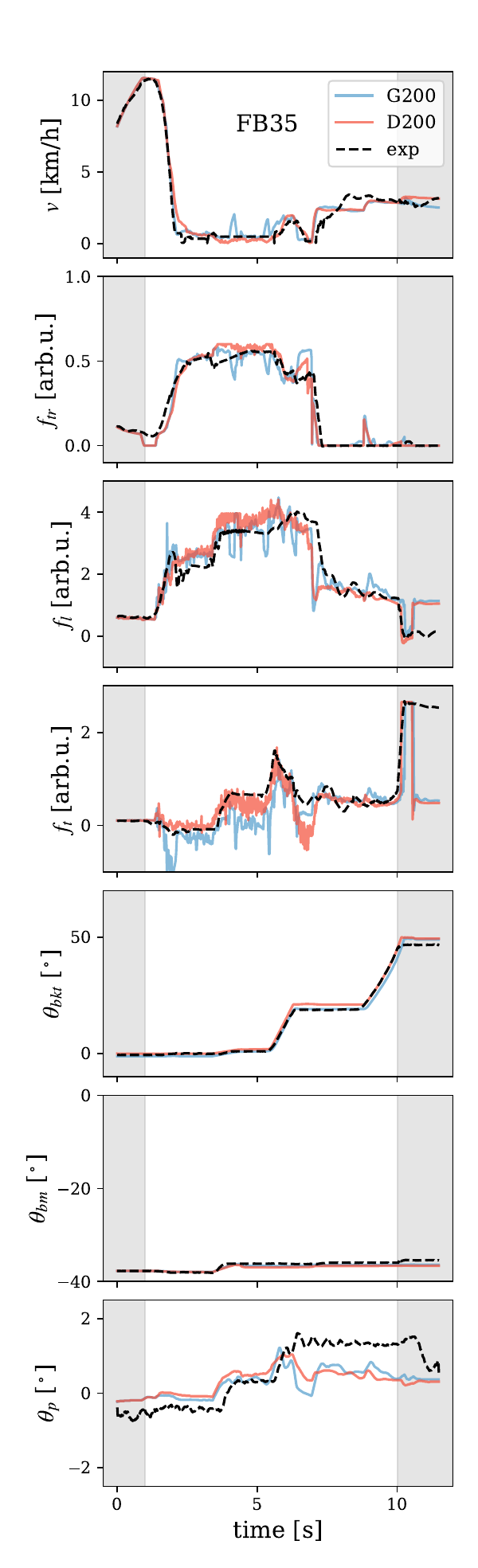} 
  \hspace{-2.5mm}
  \includegraphics[height=1.0\linewidth,trim={19mm 3mm 3mm 10mm},clip]{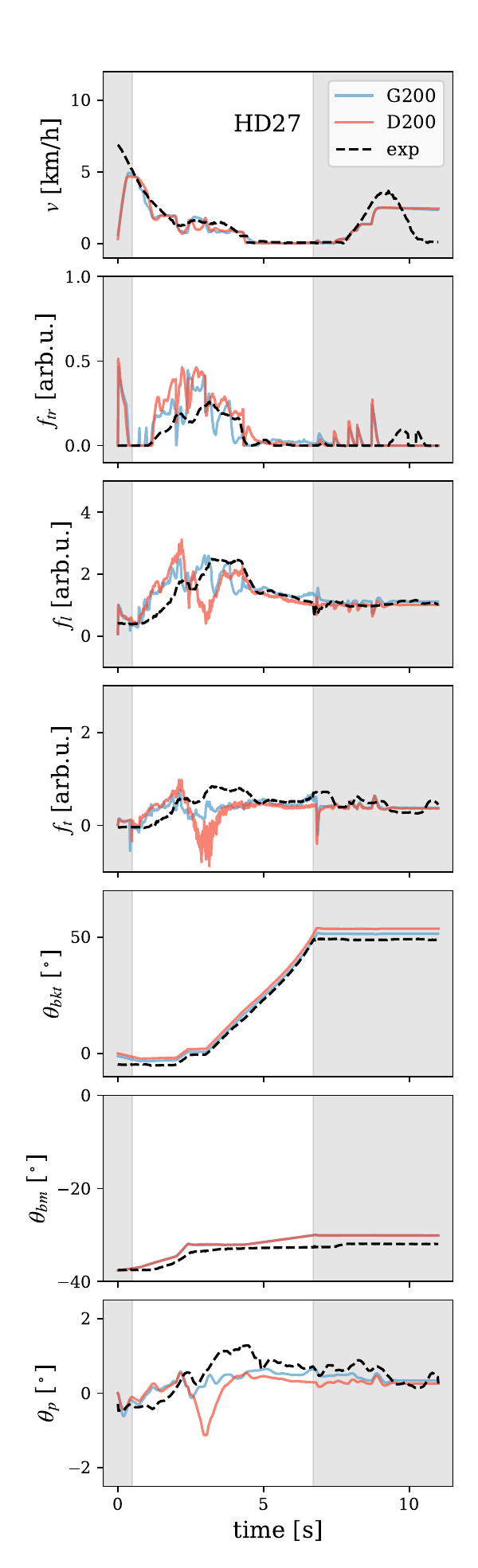}
  \hspace{-2.5mm}
  \includegraphics[height=1.0\linewidth,trim={19mm 3mm 3mm 10mm},clip]{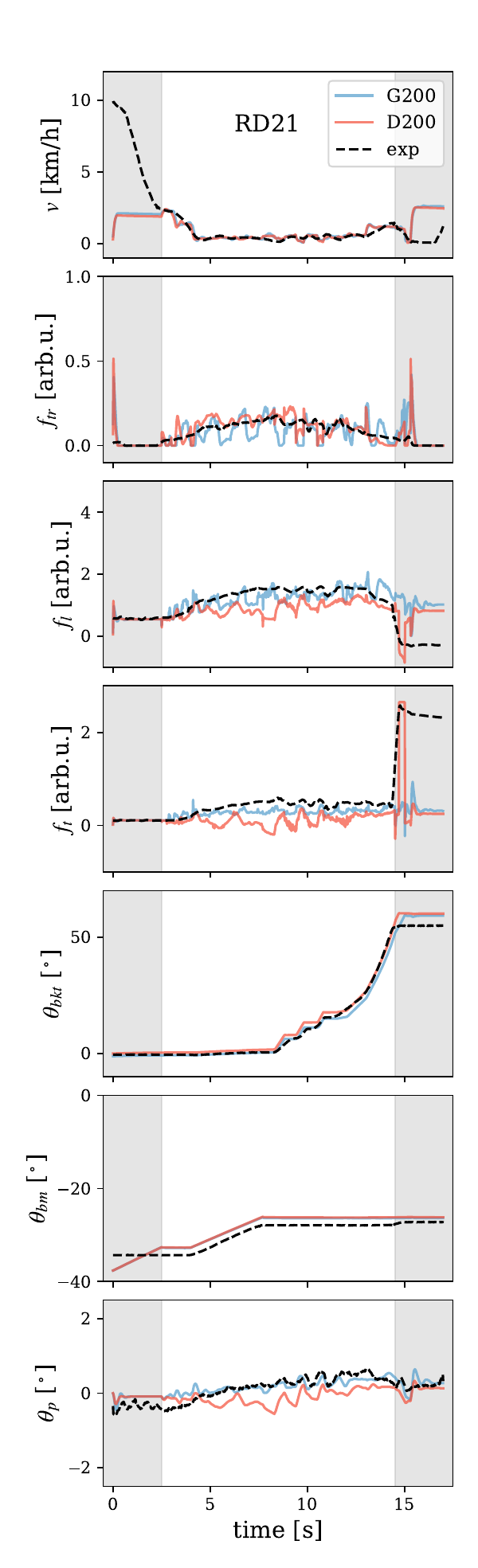}
  \caption{Speed, force, rotation measurements for time series from the FB35, HD27, RD21 experiments and type G200 and D200 simulator results.}
  \label{fig:suppl-GD200}
\end{figure}

\begin{figure}
  \centering
  \includegraphics[height=1.0\linewidth,trim={3mm 3mm 3mm 10mm},clip]{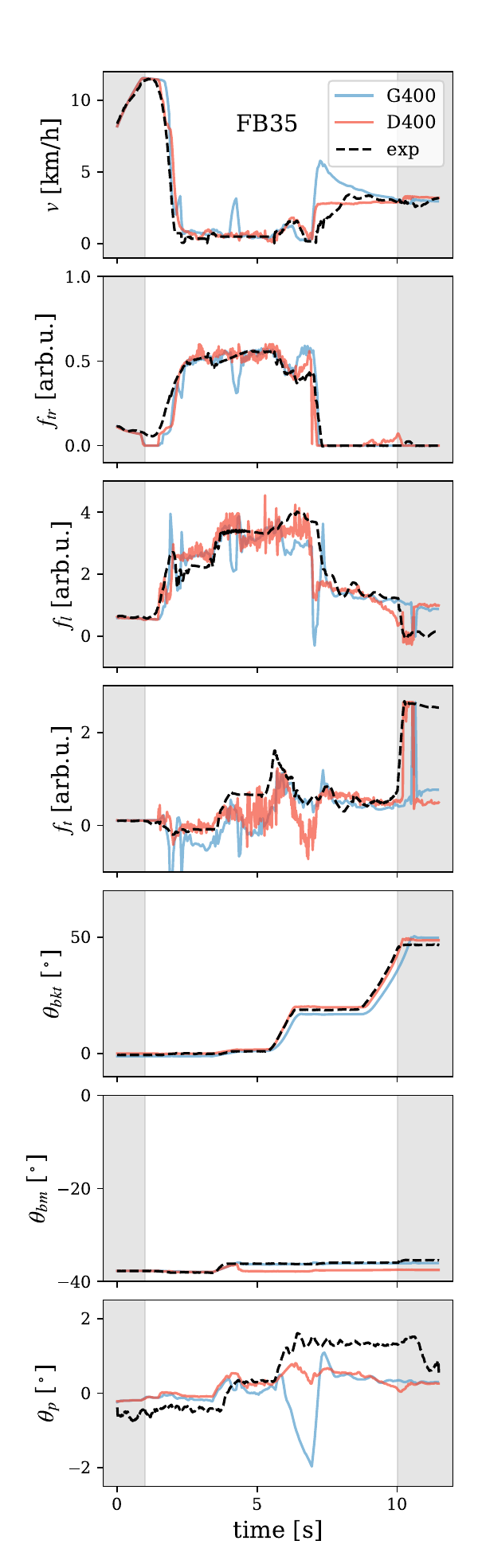} 
  \hspace{-2.5mm}
  \includegraphics[height=1.0\linewidth,trim={19mm 3mm 3mm 10mm},clip]{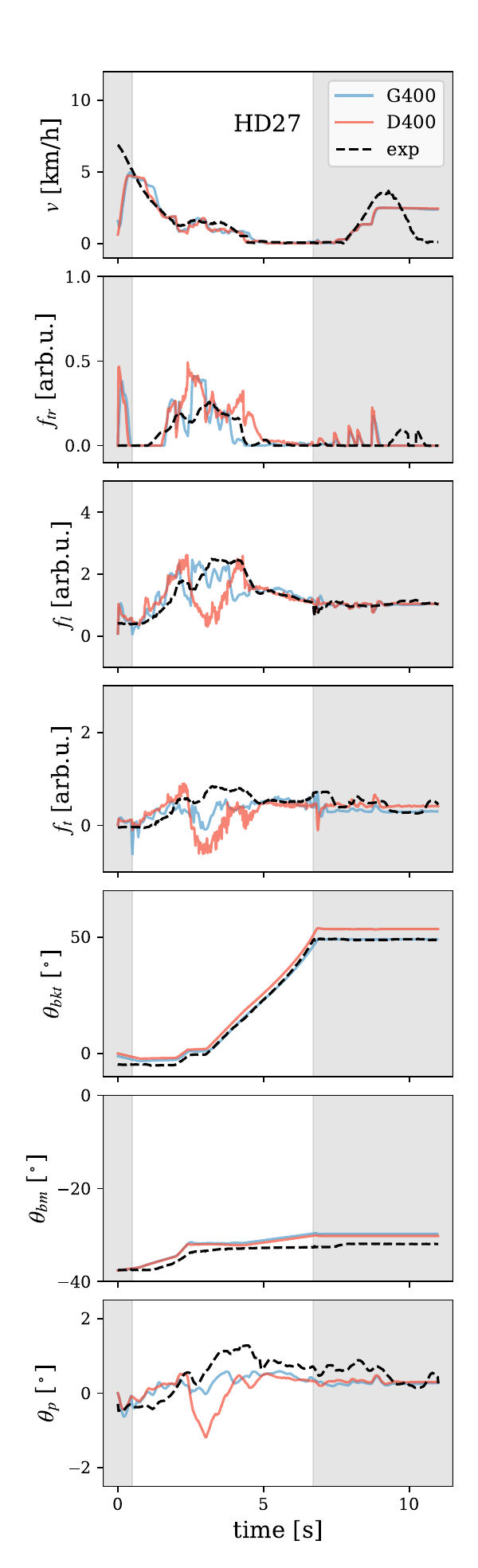}
  \hspace{-2.5mm}
  \includegraphics[height=1.0\linewidth,trim={19mm 3mm 3mm 10mm},clip]{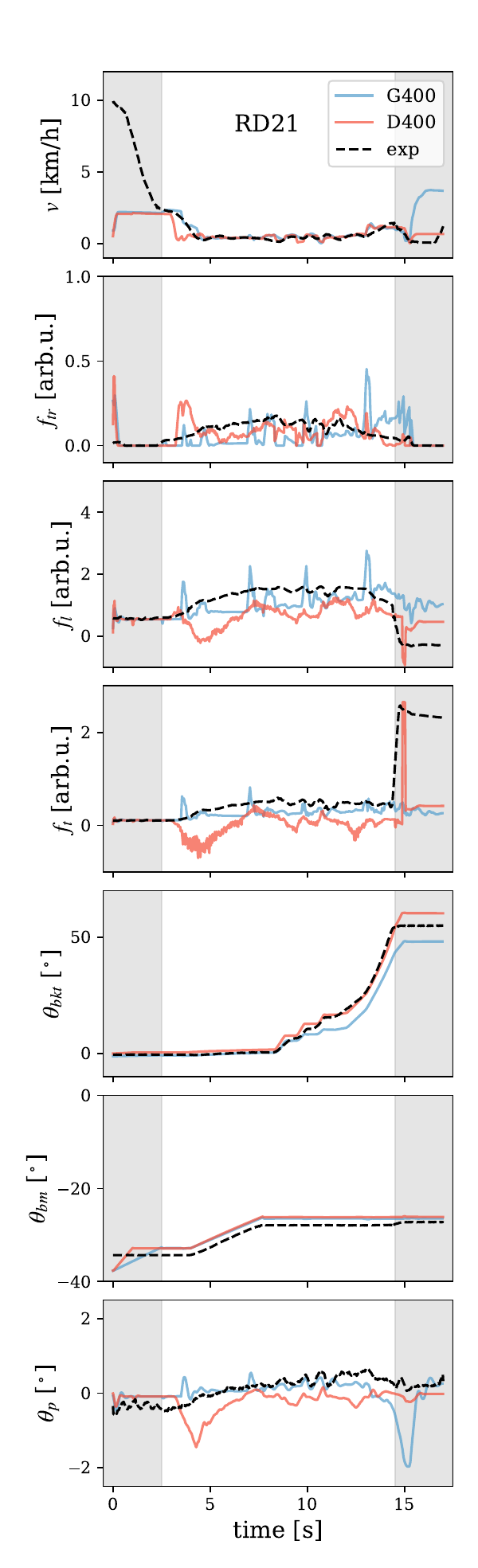}
  \caption{Speed, force, rotation measurements for time series from the FB35, HD27, RD21 experiments and type G400 and D400 simulator results.}
  \label{fig:suppl-GD400}
\end{figure}

Figure \ref{fig:force_velocity} collects measurements from all simulators in a single figure for ease of comparison but at the cost of many overlapping lines. In Fig.~\ref{fig:suppl-GD50}-\ref{fig:suppl-GD400}, we present the measurements for each spatial resolution.

\section*{Acknowledgement}
The research was supported in part by Komatsu Ltd, Algoryx Simulation AB, and the Swedish National Infrastructure for Computing at High-Performance Computing Center North (HPC2N).
\bibliographystyle{abbrv}
\bibliography{wl_sim2real_gap}

\def\cprime{$'$}
\begin{thebibliography}{10}

\bibitem{AGX20}
{AGX Dynamics}.
\newblock \url{https://www.algoryx.se/products/agx-dynamics}.
\newblock Accessed: 2023-09-01.

\bibitem{allevato2020}
A.~D. Allevato, E.~Schaertl~Short, M.~Pryor, and A.~L. Thomaz.
\newblock Iterative residual tuning for system identification and sim-to-real
  robot learning.
\newblock {\em Autonomous Robots}, 44:1167--1182, 2020.

\bibitem{aoshima:2023:pmh}
K.~Aoshima, A.~Fälldin, E.~Wadbro, and M.~Servin.
\newblock Predictor models for high-performance wheel loading.
\newblock {\em arXiv preprint arXiv:2309.12016}, 2023.

\bibitem{aoshima2021}
K.~Aoshima, M.~Servin, and E.~Wadbro.
\newblock Simulation-based optimization of high-performance wheel loading.
\newblock In C.~Feng and {et al}, editors, {\em Proceedings of the 38th
  International Symposium on Automation and Robotics in Construction (ISARC)},
  pages 688--695, Dubai, UAE, November 2021. International Association for
  Automation and Robotics in Construction (IAARC).

\bibitem{atkeson1997}
C.~G. Atkeson and S.~Schaal.
\newblock Robot learning from demonstration.
\newblock In {\em ICML}, volume~97, pages 12--20, 1997.

\bibitem{Azulay2021}
O.~{Azulay} and A.~{Shapiro}.
\newblock Wheel loader scooping controller using deep reinforcement learning.
\newblock {\em IEEE Access}, 9:24145--24154, 2021.

\bibitem{backman2021}
S.~Backman, D.~Lindmark, K.~Bodin, M.~Servin, J.~Mörk, and H.~Löfgren.
\newblock Continuous control of an underground loader using deep reinforcement
  learning.
\newblock {\em Machines}, 9(10), 2021.

\bibitem{Berndt1994}
D.~J. Berndt and J.~Clifford.
\newblock Using dynamic time warping to find patterns in time series.
\newblock In {\em Proceedings of the 3rd International Conference on Knowledge
  Discovery and Data Mining}, AAAIWS'94, page 359–370. AAAI Press, 1994.

\bibitem{choi2021use}
H.~Choi, C.~Crump, C.~Duriez, A.~Elmquist, G.~Hager, D.~Han, F.~Hearl,
  J.~Hodgins, A.~Jain, F.~Leve, et~al.
\newblock On the use of simulation in robotics: Opportunities, challenges, and
  suggestions for moving forward.
\newblock {\em Proceedings of the National Academy of Sciences},
  118(1):e1907856118, 2021.

\bibitem{collins2019}
J.~Collins, J.~McVicar, D.~Wedlock, R.~Brown, D.~Howard, and J.~Leitner.
\newblock Benchmarking simulated robotic manipulation through a real world
  dataset.
\newblock {\em IEEE Robotics and Automation Letters}, 5(1):250--257, 2019.

\bibitem{Dadhich2016}
S.~Dadhich, U.~Bodin, and U.~Andersson.
\newblock Key challenges in automation of earth-moving machines.
\newblock {\em Automation in Construction}, 68:212--222, 2016.

\bibitem{Dobson2017}
A.~Dobson, J.~Marshall, and J.~Larsson.
\newblock Admittance control for robotic loading: Design and experiments with a
  1-tonne loader and a 14-tonne load-haul-dump machine.
\newblock {\em Journal of Field Robotics}, 34(1):123--150, 2017.

\bibitem{Egli2022b}
P.~Egli and M.~Hutter.
\newblock A general approach for the automation of hydraulic excavator arms
  using reinforcement learning.
\newblock {\em IEEE Robotics and Automation Letters}, 7(2):5679--5686, 2022.

\bibitem{erez2015}
T.~Erez, Y.~Tassa, and E.~Todorov.
\newblock Simulation tools for model-based robotics: Comparison of {Bullet},
  {Havok}, {MuJoCo}, {ODE} and {PhysX}.
\newblock In {\em 2015 IEEE International Conference on Robotics and Automation
  (ICRA)}, pages 4397--4404, 2015.

\bibitem{Eriksson2023}
D.~Eriksson and R.~Ghabcheloo.
\newblock Comparison of machine learning methods for automatic bucket filling:
  An imitation learning approach.
\newblock {\em Automation in Construction}, 150:104843, 2023.

\bibitem{Filla2017}
R.~Filla and B.~Frank.
\newblock Towards finding the optimal bucket filling strategy through
  simulation.
\newblock In {\em Proceedings of 15:th Scandinavian International Conference on
  Fluid Power, June 7-9, 2017, Linköping, Sweden}, 06 2017.

\bibitem{Frank2018}
B.~Frank, J.~Kleinert, and R.~Filla.
\newblock Optimal control of wheel loader actuators in gravel applications.
\newblock {\em Automation in Construction}, 91:1--14, 2018.

\bibitem{hairer:1996:sod}
E.~Hairer and G.~Wanner.
\newblock {\em Solving Ordinary Differential Equations {II}: Stiff and
  Differential Algebraic Problems}, volume~14 of {\em Springer Series in
  Computational Mathematics}.
\newblock Springer-Verlag, Berlin, Heidelberg, New York, London, Paris, Tokyo,
  Hong Kong, second revised edition edition, 1996.

\bibitem{Holz2009}
D.~Holz, T.~Beer, and T.~Kuhlen.
\newblock Soil deformation models for real-time simulation: A hybrid approach.
\newblock In H.~Prautzsch, A.~Schmitt, J.~Bender, and M.~Teschner, editors,
  {\em Workshop in Virtual Reality Interactions and Physical Simulation
  "VRIPHYS" (2009)}, 2009.

\bibitem{Horak2019}
P.~C. Horak and J.~C. Trinkle.
\newblock On the similarities and differences among contact models in robot
  simulation.
\newblock {\em IEEE Robotics and Automation Letters}, 4(2):493--499, 2019.

\bibitem{ibarz2021}
J.~Ibarz, J.~Tan, C.~Finn, M.~Kalakrishnan, P.~Pastor, and S.~Levine.
\newblock How to train your robot with deep reinforcement learning: lessons we
  have learned.
\newblock {\em The International Journal of Robotics Research},
  40(4-5):698--721, 2021.

\bibitem{Jaiswal2019}
S.~{Jaiswal}, P.~{Korkealaakso}, R.~{Åman}, J.~{Sopanen}, and A.~{Mikkola}.
\newblock Deformable terrain model for the real-time multibody simulation of a
  tractor with a hydraulically driven front-loader.
\newblock {\em IEEE Access}, 7:172694--172708, 2019.

\bibitem{Jekel2019}
C.~F. Jekel, G.~Venter, M.~P. Venter, N.~Stander, and R.~T. Haftka.
\newblock {Similarity measures for identifying material parameters from
  hysteresis loops using inverse analysis}.
\newblock {\em International Journal of Material Forming}, may 2019.

\bibitem{Kadian2020}
A.~Kadian, J.~Truong, A.~Gokaslan, A.~Clegg, E.~Wijmans, S.~Lee, M.~Savva,
  S.~Chernova, and D.~Batra.
\newblock Sim2real predictivity: Does evaluation in simulation predict
  real-world performance?
\newblock {\em IEEE Robotics and Automation Letters}, 5(4):6670--6677, 2020.

\bibitem{Kim2016}
H.~Kim, K.~Oh, K.~Ko, P.~Kim, and K.~Yi.
\newblock Modeling, validation and energy flow analysis of a wheel loader.
\newblock {\em Journal of Mechanical Science and Technology}, 30(2):603--610,
  Feb. 2016.

\bibitem{koos2010}
S.~Koos, J.-B. Mouret, and S.~Doncieux.
\newblock Crossing the reality gap in evolutionary robotics by promoting
  transferable controllers.
\newblock In {\em Proceedings of the 12th annual conference on Genetic and
  evolutionary computation}, pages 119--126, 2010.

\bibitem{Kurinov2020}
I.~{Kurinov}, G.~{Orzechowski}, P.~{Hämäläinen}, and A.~{Mikkola}.
\newblock Automated excavator based on reinforcement learning and multibody
  system dynamics.
\newblock {\em IEEE Access}, 8:213998--214006, 2020.

\bibitem{Lacoursiere2007}
{\sc Lacoursière, C.}
\newblock {\em Ghosts and machines: regularized variational methods for
  interactive simulations of multibodies with dry frictional contacts}.
\newblock PhD thesis, Umeå University, SE-901 87 Umeå, June 2007.

\bibitem{lacoursiere2010}
C.~Lacoursi\'{e}re, M.~Linde, O.~Sabelstr{\"o}m,
\newblock Direct sparse factorization of blocked saddle point matrices,
\newblock Para 2010: State of the Art in Scientific and Parallel Computing, Reykjavik, June 6-9.2010).

\bibitem{Lindmark2018}
D.~Lindmark and M.~Servin.
\newblock Computational exploration of robotic rock loading.
\newblock {\em Robotics and Autonomous Systems}, 106:117--129, 2018.

\bibitem{Matsumoto2020}
K.~Matsumoto, A.~Yamaguchi, T.~Oka, M.~Yasumoto, S.~Hara, M.~Iida, and
  M.~Teichmann.
\newblock Simulation-based reinforcement learning approach towards construction
  machine automation.
\newblock In H.~Osumi, H.~Furuya, and K.~Tateyama, editors, {\em Proceedings of
  the 37th International Symposium on Automation and Robotics in Construction
  (ISARC)}, pages 457--464. International Association for Automation and
  Robotics in Construction (IAARC).

\bibitem{mckyes:1985:sct}
E.~McKyes.
\newblock {\em Soil cutting and tillage}.
\newblock Developments in agricultural engineering. Elsevier, Amsterdam, 1985.

\bibitem{meng2019}
Y.~Meng, H.~Fang, G.~Liang, Q.~Gu, and L.~Liu.
\newblock Bucket trajectory optimization under the automatic scooping of {LHD}.
\newblock {\em Energies}, 12(20):3919, 2019.

\bibitem{muratore2022}
F.~Muratore, F.~Ramos, G.~Turk, W.~Yu, M.~Gienger, and J.~Peters.
\newblock Robot learning from randomized simulations: A review.
\newblock {\em Frontiers in Robotics and AI}, page~31, 2022.

\bibitem{Oh2015}
K.~Oh, H.~Kim, K.~Ko, P.~Kim, and K.~Yi.
\newblock Integrated wheel loader simulation model for improving performance
  and energy flow.
\newblock {\em Automation in Construction}, 58:129--143, 2015.

\bibitem{Peng2018}
X.~B. Peng, M.~Andrychowicz, W.~Zaremba, and P.~Abbeel.
\newblock Sim-to-real transfer of robotic control with dynamics randomization.
\newblock In {\em 2018 IEEE International Conference on Robotics and Automation
  (ICRA)}, pages 1--8. IEEE Press, 2018.

\bibitem{Servin2021}
M.~Servin, T.~Berglund, and S.~Nystedt.
\newblock A multiscale model of terrain dynamics for real-time earthmoving
  simulation.
\newblock {\em Advanced Modeling and Simulation in Engineering Sciences},
  8(1):11, May 2021.

\bibitem{Servin2018}
M.~Servin and M.~Brandl.
\newblock Physics-based virtual environments for autonomous earthmoving and
  mining machinery.
\newblock In {\em Commercial Vehicle Technology Symposium – CVT 2018}, 2018.

\bibitem{servin:2014:esn}
M.~Servin, D.~Wang, C.~Lacoursi{\`e}re, and K.~Bodin.
\newblock Examining the smooth and nonsmooth discrete element approach to
  granular matter.
\newblock {\em Int. J. Numer. Meth. Engng.}, 97:878--902, 2014.

\bibitem{soderstrom1989system}
T.~S{\"o}derstr{\"o}m and P.~Stoica.
\newblock {\em System Identification}.
\newblock Prentice-Hall international series in systems and control
  engineering. Prentice Hall, 1989.

\bibitem{song2022}
R.~Song, Z.~Ye, L.~Wang, T.~He, and L.~Zhang.
\newblock Autonomous wheel loader trajectory tracking control using
  {LPV}-{MPC}.
\newblock In {\em 2022 American Control Conference (ACC)}, pages 2063--2069,
  2022.

\bibitem{tan2016}
J.~Tan, Z.~Xie, B.~Boots, and C.~K. Liu.
\newblock Simulation-based design of dynamic controllers for humanoid
  balancing.
\newblock In {\em 2016 IEEE/RSJ International Conference on Intelligent Robots
  and Systems (IROS)}, pages 2729--2736, 2016.

\bibitem{Tobin2017}
J.~{Tobin}, R.~{Fong}, A.~{Ray}, J.~{Schneider}, W.~{Zaremba}, and P.~{Abbeel}.
\newblock Domain randomization for transferring deep neural networks from
  simulation to the real world.
\newblock In {\em 2017 IEEE/RSJ International Conference on Intelligent Robots
  and Systems (IROS)}, pages 23--30, 2017.

\bibitem{Wang2022a}
S.~Wang, Y.~Yin, Y.~Wu, and L.~Hou.
\newblock Modeling and verification of an acquisition strategy for wheel
  loader's working trajectories and resistance.
\newblock {\em Sensors}, 22(16), 2022.

\bibitem{wang:2016:wsp}
D.~Wang, and M.~Servin
\newblock Warm starting the projected gauss-seidel algorithm for granular matter simulation.,
\newblock Computational Particle Mechanics. 3:43--52, 2016.

\bibitem{wiberg:2021:dem}
V.~Wiberg, M.~Servin, and T.~Nordfjell.
\newblock Discrete element modelling of large soil deformations under heavy
  vehicles.
\newblock {\em Journal of Terramechanics}, 93:11 -- 21, 2021.

\bibitem{wiberg2023}
V.~Wiberg, E.~Wallin, A.~Fälldin, T.~Semberg, M.~Rossander, E.~Wadbro, and
  M.~Servin.
\newblock Sim-to-real transfer of active suspension control using deep
  reinforcement learning.
\newblock {\em arXiv preprint arXiv:2306.11171}, 2023.

\bibitem{wiberg:2021:crt}
V.~Wiberg, E.~Wallin, T.~Nordfjell, and M.~Servin.
\newblock Control of rough terrain vehicles using deep reinforcement learning.
\newblock {\em IEEE Robotics and Automation Letters}, 7(1):390--397, 2022.

\bibitem{zhao2020sim2real}
W.~Zhao, J.~P. Queralta, and T.~Westerlund.
\newblock Sim-to-real transfer in deep reinforcement learning for robotics: a
  survey.
\newblock In {\em 2020 IEEE symposium series on computational intelligence
  (SSCI)}, pages 737--744. IEEE, 2020.

\end{thebibliography}

\end{document}